\documentclass[12pt]{report}

\usepackage[a4paper,top=1in,bottom=1in,left=1in,right=1in]{geometry}
\usepackage{amsmath, amssymb, bm}
\usepackage{caption}
\usepackage{graphicx}
\usepackage[colorinlistoftodos]{todonotes}
\usepackage[colorlinks=true, allcolors=black]{hyperref}
\usepackage{float}
\usepackage{setspace}
\usepackage{subfigure}
\usepackage{url}
\usepackage{tabularx}
\usepackage[utf8]{inputenc}
\usepackage{mathptmx} 
\usepackage{titlesec}
\titlespacing{\chapter}{0pt}{0pt}{0pt}
\usepackage{fancyhdr}
\usepackage{etoolbox}
\usepackage{appendix}
\usepackage{siunitx}
\usepackage{nomencl}

\makenomenclature
\usepackage{caption}
\usepackage{lipsum}
\usepackage{algpseudocode}
\usepackage{empheq}
\usepackage{mdwlist}
\usepackage{booktabs}
\usepackage{colortbl}
\usepackage{moresize}

\usepackage[style=ieee]{biblatex}
\DeclareCaptionType{appfigure}[Figure]
\DeclareCaptionType{apptable}[Table]

\newcommand{\chapname}{Chapter }
 
\titleformat{\chapter}[block]
{\bfseries\LARGE\centering}
{}{1em}{}[\rule{\textwidth}{0.3pt}]

\titleformat{\section}
{\bfseries\large}
{\thesection}{1em}{}

\titleformat{\subsection}
{\normalfont\bfseries}
{\thesubsection}{1em}{}

\titleformat{\subsubsection}
{\normalfont\bfseries}
{\thesubsubsection}{1em}{}

\usepackage{fancyhdr}
\usepackage{multirow}
\usepackage{multicol}

\usepackage{placeins}

\makeatletter
\renewcommand*\env@matrix[1][*\c@MaxMatrixCols c]{%
  \hskip -\arraycolsep
  \let\@ifnextchar\new@ifnextchar
  \array{#1}}
\makeatother

\usepackage{enumitem}

\usepackage{listings}
\usepackage{color}
 
\definecolor{codegreen}{rgb}{0,0.6,0}
\definecolor{codegray}{rgb}{0.5,0.5,0.5}
\definecolor{codepurple}{rgb}{0.58,0,0.82}
\definecolor{backcolour}{rgb}{0.95,0.95,0.95}

\newcommand{\codesize}{\fontsize{10pt}{11pt}\selectfont}

\lstdefinestyle{mystyle}{
    backgroundcolor=\color{backcolour},   
    commentstyle=\color{codegreen},
    keywordstyle=\color{magenta},
    numberstyle=\tiny\color{codegray},
    stringstyle=\color{codepurple},
    basicstyle=\ttfamily\codesize,
    breakatwhitespace=true,         
    breaklines=true,                 
    captionpos=b,                    
    keepspaces=false,                 
    numbers=left,                    
    numbersep=5pt,                  
    showspaces=false,                
    showstringspaces=false,
    showtabs=false,                  
    tabsize=2,
    showlines = true,
    fontadjust = true,
    framexleftmargin = 10 pt,
    resetmargins = true,
    basewidth = 0.5em
}
 

\usepackage{multicol}
 
\usepackage{makecell}
\usepackage{emptypage}

\newcolumntype{R}{>{\raggedleft\arraybackslash}X}
\newcolumntype{L}{>{\raggedright\arraybackslash}X}
\newcolumntype{C}{>{\centering\arraybackslash}X}

\usepackage{textcomp}
\usepackage{subcaption}
\usepackage{setspace}
\renewcommand{\baselinestretch}{1.5}
\begin{document}
\sloppy
\begin{titlepage}
\begin{figure}[!t]
\centering
\includegraphics[width = 4.2in]{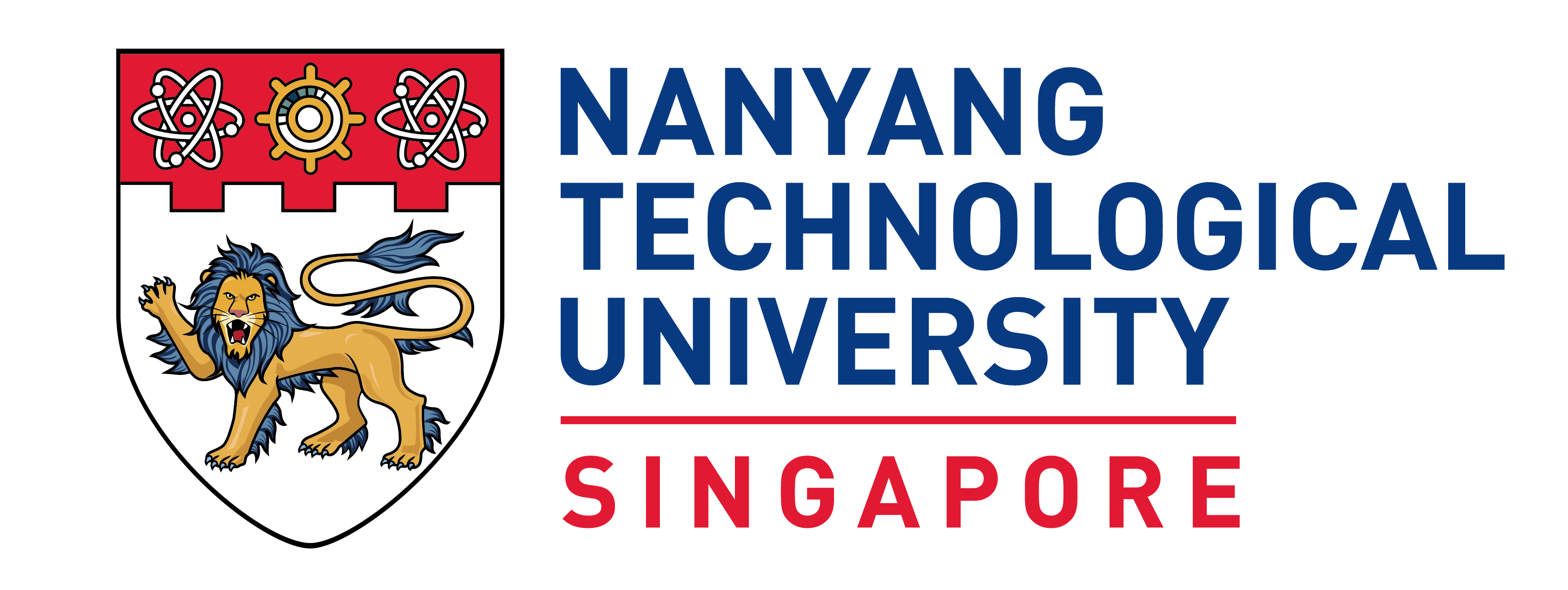}
\caption*{}
\end{figure}

\centering
\LARGE{\textbf{A Low-cost and Portable Active Noise Control Unit}}\\[1in]

\LARGE{\textbf{Wang Zhaohan}}\\[0.2in]
\normalsize{Matriculation Number: U1822545B}\\[1in]


\large{\textbf{School of Electrical \& Electronic Engineering}} \\[1in]

\large{A final year project report presented to the Nanyang Technological University
in partial fulfilment of the requirements of the degree of
Bachelor of Engineering }\\[1in]


\Large{2022}
\newpage
\end{titlepage}

\pagenumbering{roman}
\pagestyle{fancy}
\fancyhf{}
\cfoot{\thepage}


\chapter*{Abstract}
\rhead{Abstract}
\addcontentsline{toc}{chapter}{Abstract}
The objective of this research is to employ cutting-edge active noise control methodologies in order to mitigate the noise emissions produced by electrical appliances, such as a coffee machine. The algorithm utilized in this study is the modified Filtered-X Least Mean Square (FXLMS) algorithm. This algorithm aims to generate an anti-noise waveform by utilizing measurements from both the reference microphone and the error microphone. The desired outcome of this approach is to achieve a residual noise level of zero. The primary difficulty lies in conducting the experiment in an open space setting, as conventional active noise control systems are designed to function within enclosed environments, such as closed rooms or relatively confined spaces like the volume inside headphones. A validation test bench is established, employing the Sigma Studio software to oversee the entire system, with the ADAU1452 digital signal processor being chosen. This study presents an introduction to different Active Noise Control systems and algorithms, followed by the execution of simulations for representative techniques. Subsequently, this section provides a comprehensive account of the procedures involved in executing the experiments, followed by an exploration of potential avenues for further research.
\\
\par
\textbf{Keywords:} Active noise control, Filtered reference least mean square algorithm (FXLMS), Low-cost active noise control system. 


\newpage



\chapter*{Acknowledgement}
\rhead{Acknowledgement}
\addcontentsline{toc}{chapter}{Acknowledgement}
I am writing to extend my sincere appreciation to Gan Woon Seng, my supervising professor, for affording me the invaluable opportunity to complete the project under his guidance. He delineated the pertinent principles and assigned me duties with lucidity and accuracy. His unwavering accessibility during challenging situations served as a model for researchers and exemplified professionalism.

Additionally, I wish to express my gratitude to Assistant Professor Shi Dongyuan for his patient guidance throughout my theoretical studies and experiments. His account of his personal experience as an undergraduate student expediting the initiation of experiments and maintaining progress is truly remarkable.


\newpage
\setcounter{tocdepth}{2}

\tableofcontents
\rhead{Table of Contents}
\newpage


\renewcommand{\listfigurename}{Lists of Figures}
\rhead{Lists of Figures}
\listoffigures 
\addcontentsline{toc}{chapter}{Lists of Figures}

\newpage

\listoftables 
\addcontentsline{toc}{chapter}{Lists of Tables}
\rhead{Lists of Tables}
\newpage

\rhead{}

\pagenumbering{arabic}
\lhead{Introduction}

\chapter{1. Introduction}

The objective of a conventional Active Noise Control system is to counteract disturbances in the acoustic system in order to eliminate sounds originating from primary sources~\cite{elliott1993active,lam2021ten,shi2023active}. Active noise control is distinguished from passive noise control methods, such as sound walls and absorbers (e.g., sponges), which can be combined for improved results. Activation of noise control would be uneconomical if the frequencies to be controlled exceeded 500 Hertz, which is the norm in practical situations. Nevertheless, this frequency range encompasses most of the white sounds in our everyday environment. Active noise control is predicated on the advancement of digital signal processing techniques and electrical controllers~\cite{kuo1996design,shi2016comparison}. Consequently, the introduction of signal processing algorithms and electrical components is imminent; for instance, Burgess invented the first adaptive filtering system subsequent to the development of the digital signal processor~\cite{burgess1981active}. Furthermore, the experimental system incorporates an Analog Devices processing unit comprising high-performance capacitors, filters, amplifiers, and other components. Other electrical devices utilized include microphones, speakers, and high-speed signal cables. Subsequently, the primary limitations of the ANC system can be identified: initially, the challenge of mitigating complex noise sources~\cite{moazzam2014performance}, which arises from the algorithms themselves~\cite{shi2017effect,shi2021optimal}; strategies to enhance the algorithms will be elaborated upon in the following section; and second, the system's robustness, as the counter noise occasionally becomes an additional noise source. Active noise control systems can be classified into two categories: waveform synthesis and adaptive filtering. Adaptive filtering is a technique that generates a counter-sound wave that is in continuous motion in order to counteract the corresponding incoming noise. In contrast, waveform synthesis involves the generation of periodic waveforms~\cite{qiu2002waveform}. As a result of the limitation that waveform synthesis can only effectively handle periodic noises with a narrow bandwidth, adaptive filtering has gained widespread acceptance across various professions and industries. A few well-known examples of this limitation include noise control systems in aircraft cabins, windows~\cite{lam2020active,shi2017algorithms,shi2016open,hasegawa2018window,shi2020feedforward,lam2023anti}, earphones~\cite{shen2022adaptive,shen2022multi,shen2021alternative,shen2023implementations,shen2021wireless}, and air conditioning units~\cite{fatima2012noise,elliot1990flight,lam2020active,lam2018active}, since it can be applied in different scenarios and has a stronger capability, and as expressed by its name, adaptive system is agile in adjusting to changes in noise source, thus they are not likely to become ineffective. Adaptive filtering can be accomplished via feedforward control, feedback control, or hybrid control~\cite{lee2022compact,shi2020algorithms}, which is also utilized in practice. The following is a compilation of common single-channel and multichannel ANC systems.

\subsection{Single-channel feedforward ANC}

\begin{figure}[!t]
    \centering
    \includegraphics[width = 12cm]{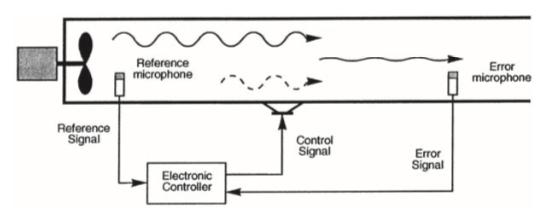}
    \caption{Single-channel feedforward ANC system in an air duct~\cite{hansen1999understanding}.}
    \label{fig:feedforward}
\end{figure}

The block diagram of a single-channel feedforward active noise control (ANC) system in an air duct is depicted in Figure~\ref{fig:feedforward}. The single-channel active noise control (ANC) system consists of a pair of microphones, a loudspeaker, and an electrical controller. The microphone positioned at the upstream of the duct is the previously mentioned reference microphone, while the microphone situated at the downstream serves as the error microphone. The reference signal $x(n)$, which is obtained by sampling the primary noise, is inputted into the controller to generate the control signal $y(n)$ that is emitted by the secondary source. Ideally, the control signal is sent through the secondary way, also known as the anti-noise path, with the objective of completely nullifying the primary noise at the precise location of the faulty microphone. Simultaneously, the error microphone detects the residual error signal, which is then utilized to modify the control signal of the controller in order to attain optimal noise reduction. Therefore, utilizing the reference microphone and the error microphone, the anti-noise produced may effectively adapt to changes in the primary noise, enabling it to effectively mitigate various types of noise, including low-frequency broadband noise. Hence, feedforward active noise control (ANC) is frequently employed in diverse industrial sectors, such as the reduction of noise in ducts~\cite{kuo1996active}.



\section{Single channel feedback ANC}

\begin{figure}[!t]
    \centering
    \includegraphics[width = 12cm]{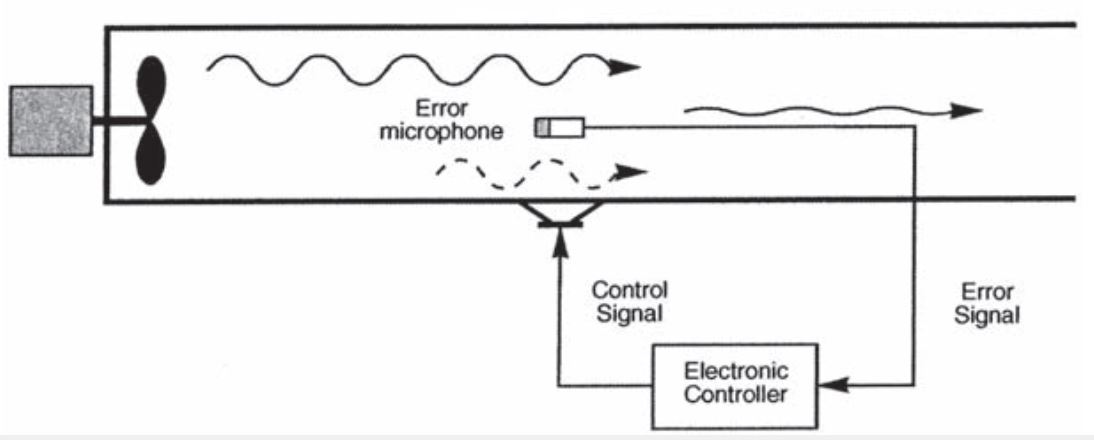}
    \caption{Single-channel feedback ANC system in an air duct~\cite{hansen1999understanding}.}
    \label{fig:feedback}
\end{figure}

Another often-used ANC implementation is the single-channel feedback ANC system, as depicted in Figure~\ref{fig:feedback}. In contrast to feedforward active noise control (ANC), feedback ANC alone necessitates the employment of a single microphone to monitor the residual error. Therefore, it is necessary for the adaptive noise cancellation (ANC) system to accurately estimate or forecast the characteristics of the primary noise that is present in the residual error. The prediction of random or broadband signals is more challenging compared to periodic signals due to the constraints imposed by system causality. Hence, it should be noted that active noise control (ANC) is mostly effective in addressing narrow-band noise and periodic noise, as stated by Kuo in his work~\cite{kuo1996active}. Nevertheless, the straightforward composition and condensed dimensions of the device offer the benefit of convenient integration into portable commercial apparatus, exemplified by the application of ANC headphones~\cite{shen2021implementation}.

\section{Multichannel Active Noise Control system}

\begin{figure}[!t]
    \centering
    \includegraphics[width = 12cm]{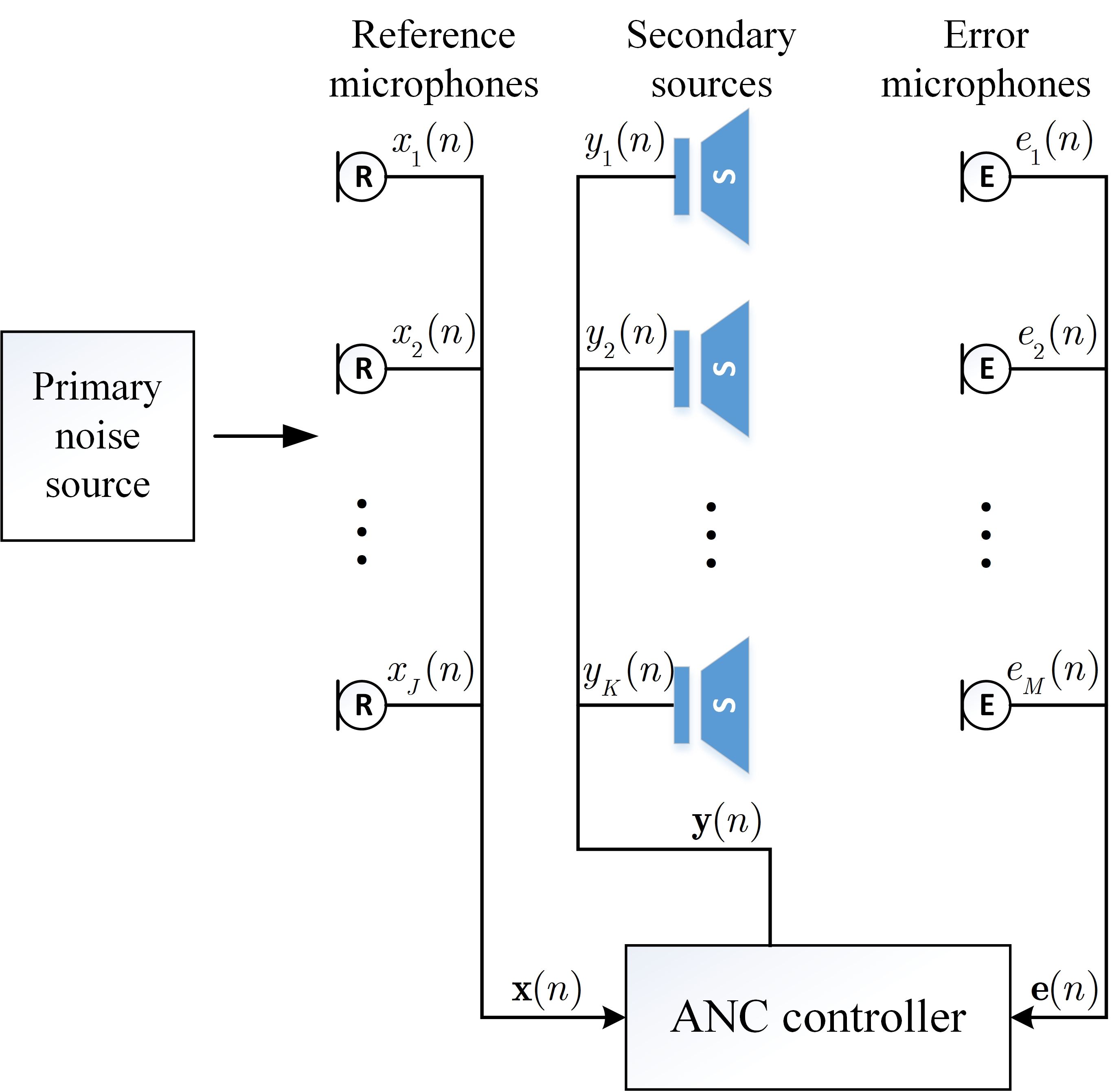}
    \caption{Block diagram of the MCANC system~\cite{shi2020algorithms}.}
    \label{fig:MCANC}
\end{figure}

The ANC configurations depicted in Figure~\ref{fig:feedforward} and Figure~\ref{fig:feedback} are examples of single-channel structures commonly employed for local noise control. These setups typically create a quiet zone around the error microphone in a free-field environment. In the context of vehicle noise cancellation, it is worth noting that the implementation of single-channel active noise control (ANC) is limited in its ability to effectively attenuate undesired sounds throughout the entire car cabin. Instead, it primarily focuses on generating quiet zones, specifically around the headrest area of the vehicle. On the other hand, the global method necessitates the reduction of a substantial open area, such as an entire room or a large duct. In this particular scenario, the utilization of multichannel active noise control (ANC) is employed with the objective of achieving a larger quiet zone by means of several secondary sources and error microphones~\cite{shi2017multiple,shi2017understanding,shi2019practical,he2019exploiting}. The figure presented in Figure~\ref{fig:MCANC} illustrates the block diagram of a multichannel active noise control (ANC) system. The multichannel active noise control (MCANC) system consists of $J$ reference microphones, $K$ secondary sources, and $M$ error microphones, where $J$ is more than zero, $K$ is greater than zero, and $M$ is greater than zero. 

\section{Tips for optimizing ANC systems}
In the realm of ANC systems, the performance is significantly influenced by both physical design and algorithm design. However, it is crucial to prioritize understanding the specific noise that the system encounters as the fundamental groundwork for subsequent endeavors. Accurate simulation can be achieved by the utilization of computer-generated noise, while the choice of anti-noise output speaker is contingent upon the specific environmental conditions and noise attributes. 

Consider the feedforward system as a case in point, which is employed in our subsequent experimental description. In the realm of physical design, there exist three noteworthy potential enhancements that can be implemented to enhance the performance of the system. These include the strategic positioning of reference microphones, the optimal placement of error microphones, and the arrangement of the control source. The optimal placement of the anti-noise speakers, as stated in the theorem, is obtained by quadratic optimization. The ultimate position is identified as the location where the total sound pressure from all error sensors is minimized. If the objective is to decrease wideband noise, additional control sources can be strategically arranged to enhance noise reduction, if deemed necessary. In practical investigations, trial and error methods are often employed due to the fact that the ideal position may vary for each noise frequency. In addition, it is imperative to explore alternative system layouts or structures, such as the inclusion of additional channels or control sources. 

Additionally, it is imperative to ensure that the error sensors are capable of detecting all sound signals generated by both the primary and secondary sources within the entirety of the experimental setting. In theory, the optimal position is determined by the maximum disparity in sound pressure between the primary sound field and the secondary sound field. Nevertheless, in cases when the theoretically optimal placement is situated at a considerable distance from the sources of noise, it becomes necessary to make adjustments. Empirically, it has been found that placing error sensors in close proximity to the control sources yields superior results.

In our experiment, it is necessary to consider the delay of the reference signal sensor when applying a feedforward system. Due to the inherent properties of electrical components, such as the analog to digital interface and the control source's loudspeaker, a significant delay of three to ten milliseconds typically occurs between the input of the reference signal and the output of the anti-noise signal. This delay necessitates compensation through either physical distance or algorithmic methods. Furthermore, the issue of reference signal quality is of paramount importance. Notably, careful attention is devoted to the coherence between signals obtained from the reference microphone and error signals, as well as the spectral characteristics of the noise source, specifically the primary frequency caught by the reference microphone. In order to achieve coherence, the optimum noise control for a given noise segment, denoted as the error signal, can be formally defined as follows:
\begin{equation}
    \frac{S_{ee}(\omega)_\mathrm{opt}}{S_{ee}(\omega)_\mathrm{unc}}=1-\gamma^2(\omega).
\end{equation}
To provide clarification, the abbreviation ``unc'' represents the spectrum of the uncontrolled signal, whereas ``opt'' denotes the spectrum of the managed signal. The symbol $\gamma$ represents the coherence coefficient between the reference signal and the error signal. A value of $\gamma$ closer to $1$ indicates a higher likelihood of the system achieving significant noise reduction. 

\newpage

\lhead{ANC algorithms and respective simulations}
\chapter{2. Active noise control algorithms and respective simulations}
Feedforward active noise control (ANC) has proven to be effective in mitigating broadband noise, making it a widely adopted technique in various applications. Nevertheless, the implementation of feedforward active noise control (ANC) using analog circuits sometimes presents challenges and difficulties. In contrast, the utilization of adaptive filter algorithms is of utmost importance in the present feedforward active noise reduction technique due to their exceptional performance and convenient implementation in digital processors. This chapter provides a concise overview of the fundamental idea behind the theory of adaptive filtering, as well as an examination of some adaptive algorithms commonly employed in the context of active noise reduction applications.

\section{Adaptive filter theory}

The application of active noise control (ANC) technology faced significant challenges, leading to a prolonged period of stagnation, until the introduction of the adaptive filter by Widrow in 1975~\cite{widrow1975adaptive}. Throughout this period, numerous researchers have acknowledged the utilization of analog circuits in the signal-channel ANC system. However, the system's restricted noise reduction capabilities and inherent instability have failed to sufficiently persuade the general public about the practicality and feasibility of ANC systems~\cite{hansen1999understanding}. Nevertheless, due to the swift advancement of adaptive filtering theory~\cite{farhang2013adaptive,haykin2002adaptive,li2013frequency,shi2016adaptive}, numerous adaptive control structures have been suggested and implemented in the field of active noise control (ANC)~\cite{kuo1996active}. In addition, the utilization of robust microprocessors in commercial settings has expedited the implementation of compact active noise control (ANC) systems~\cite{nelson1991active,george2013advances}. 

\begin{figure}[!t]
  \centering
    \includegraphics[scale=1.3]{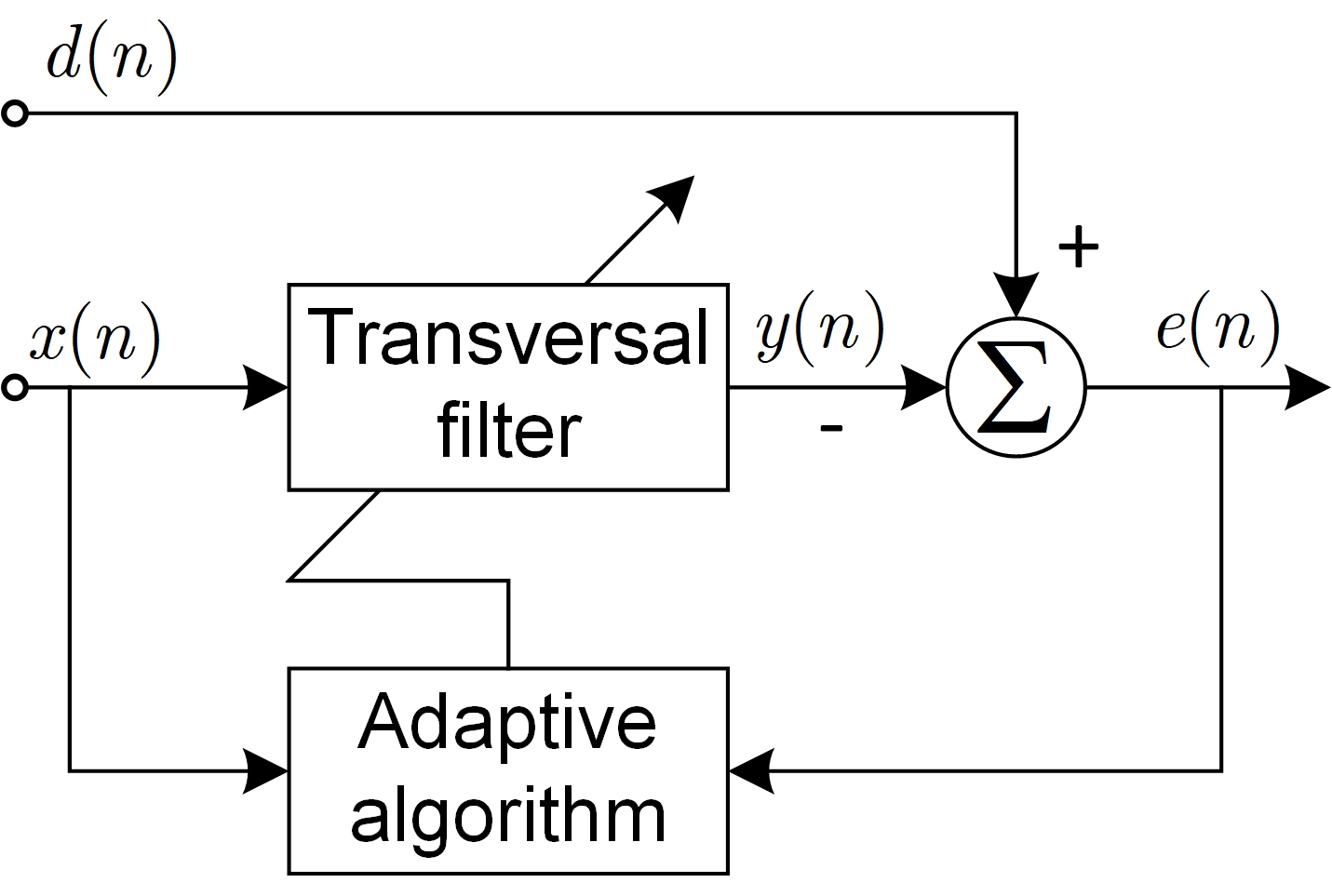}
  \caption{Block diagram of adaptive filter algorithm.}
  \label{Fig2_1}
\end{figure}

Subsequently, the adaptive filter emerged as the central component of the contemporary Active Noise Control (ANC) technique due to its notable attributes, including its resilience, superior noise control capabilities, and straightforward implementation. In order to provide a concise overview of the adaptive algorithm, we will examine an adaptive filter within the scope of this section, as depicted in Figure~\ref{Fig2_1}. The given diagram illustrates the utilization of a transversal filter to generate the signal $y(n)$ with the purpose of mitigating the presence of noise $d(n)$. 

The transversal filter's block diagram is depicted in Figure 2.2. The filter's input signal vector is represented by 
\begin{equation}\label{eq2_1}
    \mathbf{x}(n)=\left[x(n),x(n-1),\cdots,x(n-N+1)\right]^\mathrm{T}
\end{equation}
Moreover, the weight vector of the filter is written as 
\begin{equation}\label{eq2_2}
    \mathbf{w}(n)=\left[w_0(n),w_1(n),\cdots,w_\mathrm{N-1}(n)\right]^\mathrm{T}
\end{equation}
where $[\cdot]^\text{T}$ denotes the transpose operation. Hence, the output signal of the filter is obtained as 
\begin{equation}\label{eq2_3}
    y(n)=\mathbf{w}^\mathrm{T}(n)\mathbf{x}(n).
\end{equation}

The residual error of the adaptive filter can be expressed as 
\begin{equation}\label{eq2_4}
    \begin{split}
        e(n) &= d(n)-y(n)= d(n)-\mathbf{w}^\mathrm{T}(n)\mathbf{x}(n).
    \end{split}
\end{equation}
The cost function $J(n)$ is the mean square error (MSE) of (\ref{eq2_4}) given by 
\begin{equation}\label{eq2_5}
    \begin{split}
        J(n) &= \mathbb{E}\left[e^2(n)\right]\\
             &= \mathbb{E}\left[d^2(n)\right]-2\mathbf{p}^\mathrm{T}\mathbf{w}(n)+\mathbf{w}^\mathrm{T}(n)\mathbf{R}\mathbf{w}(n)
    \end{split}
\end{equation}
where $\mathbb{E}\left[\cdot\right]$ denotes the expectation operator. $\mathbf{p}$ denotes the cross-correlation vector given by 
\begin{equation}\label{eq2_6}
    \mathbf{p}=\mathbb{E}\left[d(n)\mathbf{x}(n)\right]
\end{equation}
and $\mathbf{R}$ represents the auto-correlation matrix of the input signal as 
\begin{equation}\label{eq2_7}
    \mathbf{R} = \mathbb{E}\left[\mathbf{x}(n)\mathbf{x}^\mathrm{T}(n)\right].
\end{equation}
The optimal filter is obtained by minimizing (\ref{eq2_5}), where the gradient of (\ref{eq2_5})is set to zero, giving 
\begin{equation}\label{eq2_8}
    \mathbf{w}_\mathrm{o}=\mathbf{R}^\mathrm{-1}\mathbf{p}
\end{equation}
where $\mathbf{R}$ is assumed to be invertible. Substituting (\ref{eq2_8}) into (\ref{eq2_5}) yields the minimum mean square error (MMSE)
\begin{equation}\label{eq2_9}
    J_\mathrm{min}=\mathbb{E}\left[d^2(n)\right]-\mathbf{p}^\mathrm{T}\mathbf{w}_\mathrm{o}.
\end{equation}
Combined with (\ref{eq2_9}), (\ref{eq2_5}) is rewritten as 
\begin{equation}\label{eq2_10}
    \begin{split}
        J(n) &= J_\mathrm{min} +\left[\mathbf{w}(n)-\mathbf{w}_\mathrm{o}\right]^\mathrm{T}\mathbf{R}\left[\mathbf{w}(n)-\mathbf{w}_\mathrm{o}\right]\\
        &= J_\mathrm{min} + \mathbf{v}^\mathrm{T}\mathbf{R}\mathbf{v}
    \end{split}
\end{equation}
where
\begin{equation}
    \mathbf{v} = \mathbf{w}(n)-\mathbf{w}_\mathrm{o}.
\end{equation}
\begin{figure}[!t]
  \centering
    \includegraphics[scale=1.2]{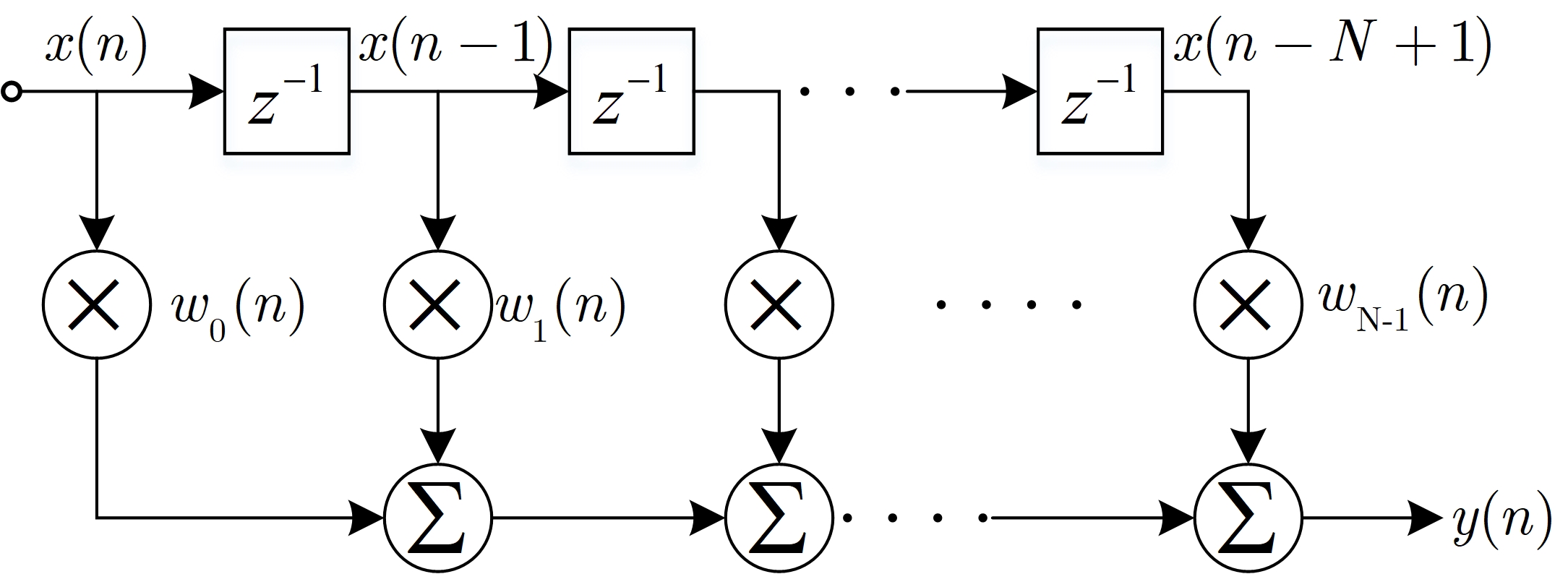}
  \caption{Block diagram of the transversal filter.}
  \label{Fig2_2}
\end{figure}

\section{Least mean square (LMS) algorithm}
\label{Chapter_2_LMS}
The preceding section demonstrates the most effective solution of the adaptive algorithm. However, the utilization of equation (\ref{eq2_8}) for attaining the best filter is not feasible in practical scenarios, as the cross-correlation vector (\ref{eq2_6}) and the auto-correlation matrix (\ref{eq2_7}) are typically not accessible. In order to address this matter, we substitute the mean square error (\ref{eq2_5}) with its instantaneous value:
\begin{equation}\label{eq2_12}
    J(n) =  e^2(n).
\end{equation}
The gradient of (\ref{eq2_12}) with respect to $\mathbf{w}(n)$ is given by 
\begin{equation}\label{eq2_13}
    \nabla J(n) = -2\mathbf{x}(n)e(n).
\end{equation}
Therefore, the recursive formula for the filter weight is derived to
\begin{equation}\label{eq2_14}
    \mathbf{w}(n+1) = \mathbf{w}(n)+\mu\mathbf{x}(n)e(n)
\end{equation}
where $\mu$ denotes the step size, which is a positive value and used to control the update rate of the coefficients. Equation~(\ref{eq2_14}) is the so-called least mean square error (LMS) algorithm. Compared to other adaptive algorithms, the LMS filter is most widely implemented due to its relatively low computational load and effectiveness. 

\subsection{Simulation of the LMS algorithm}
\begin{figure}[!t]
    \centering
    \includegraphics[width=12cm]{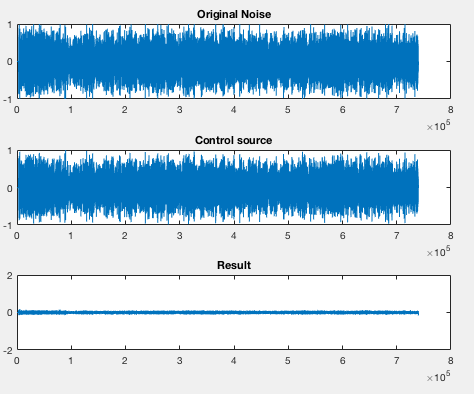}
    \caption{Simulation result of the LMS algorithm}
    \label{fig:Picture11}
\end{figure}
\begin{figure}[!t]
    \centering
    \includegraphics[width=12cm]{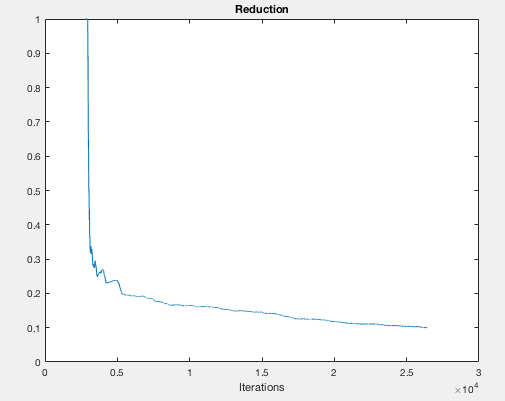}
    \caption{Learning curve}
    \label{fig:Picture12}
\end{figure}
The noise sample utilized for the MATLAB simulation, referred to as 'sample.m4a', is characterized as white noise, with a frequency range spanning from 0Hz to 1500Hz. The MATLAB code is included in the Appendix section for reference purposes. In general, theoretical simulations provide optimal outcomes. The spectra of the original noise and control source, as depicted in Figure~\ref{fig:Picture11} and \ref{fig:Picture12}, exhibit similarities. Additionally, the residual noise is observed to be consistently low and stable.

\section{Filtered-x least mean square (FxLMS) algorithm}
\label{Chapter_2_FxLMS}
The filtered-x least mean square (FxLMS) algorithm is widely regarded as one of the most effective adaptive algorithms for mitigating the effects of the secondary path in an active noise control (ANC) system~\cite{shi2016systolic}. This section provides an introduction to the FxLMS algorithm, which is utilized in feedforward single- and multichannel active noise control (ANC) applications.

\subsection{Single-channel FxLMS algorithm}
The LMS filter is commonly employed due to its efficient processing capabilities. However, its application in the ANC system is hindered by the significant deterioration of convergence behavior and stability caused by the secondary path. The proposed approach in this scenario involves the utilization of the filtered-x least mean square (FxLMS) method to mitigate the impact of the secondary path, while simultaneously maintaining a low level of computational complexity. The block diagram depicting the FxLMS algorithm is presented in Figure 2.3. The given diagram illustrates the representation of various components in the system. The transfer function of the primary path, denoted as $P(z)$, describes the relationship between the reference microphone and the error microphone. The control filter is represented by $W(z)$, while the transfer function of the secondary path from the secondary source to the error microphone is denoted as $S(z)$. The estimation of this secondary path is denoted as $\hat{S}(z)$. 
\begin{figure}[htbp]
  \centering
    \includegraphics[scale=0.9]{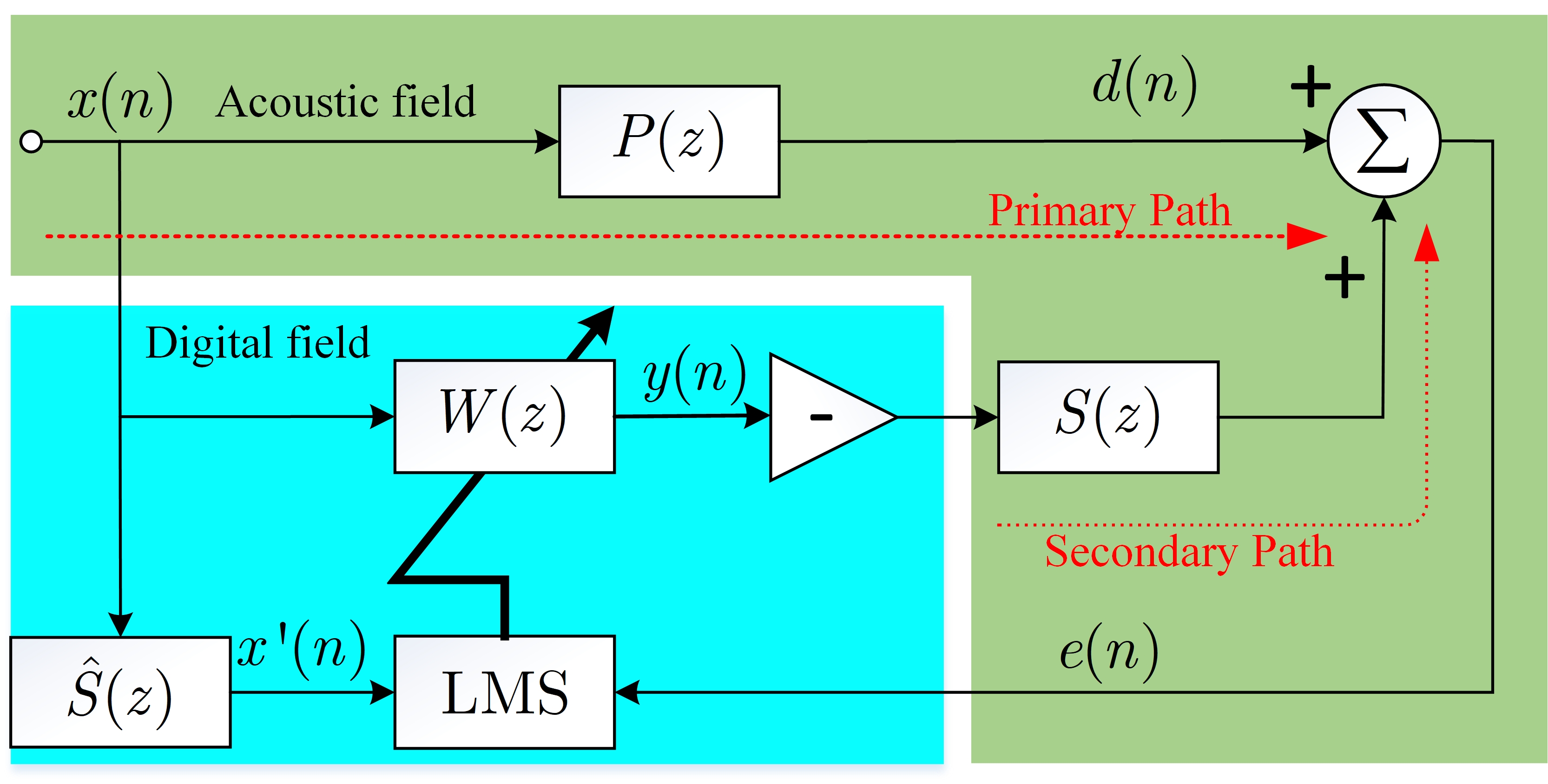}
  \caption{The diagram illustrates the implementation of the FxLMS algorithm within a single-channel feedforward active noise control (ANC) system. (1) The acoustic field is represented by the color green, whereas (2) the digital field is represented by the color blue.}
  \label{Fig2_3}
\end{figure}

The reference signal vector $\mathbf{x}(n)$ is $\left[x(n),x(n-1),\cdots,x(n-N+1)\right]^\mathrm{T}$ and $N$ is the length of the control filter. So, the output of the control filter is
\begin{equation}\label{eq2_15}
    y(n)=\mathbf{w}^\mathrm{T}(n)\mathbf{x}(n)
\end{equation}
where, $\mathbf{w}(n)$ represents the coefficient vector of the control filter. The residual error signal can be expressed as 
\begin{equation}\label{eq2_16}
    e(n) = d(n)-s(n)\ast\left[\mathbf{w}^\mathrm{T}(n)\mathbf{x}(n)\right].
\end{equation}
The primary disturbance and the impulse response of the secondary path are represented by $d(n)$ and $s(n)$, respectively. The linear convolution is denoted by $\ast$. In order to minimize the instantaneous squared error, denoted as $J(n)=e^2(n)$, the gradient descent approach is commonly employed. This method involves updating the coefficient vector in the direction opposite to the gradient, with a step size denoted as $\mu$:
\begin{equation}\label{eq2_17}
    \mathbf{w}(n+1)=\mathbf{w}(n)-\frac{\mu}{2}\nabla J(n)
\end{equation}
where $\nabla J(n)$ denotes the instantaneous estimate of the MSE gradient at time $n$, and can be expressed as 
\begin{equation}\label{eq2_18}
    \nabla J(n)=\nabla e^2(n)=-2\mathbf{x}'(n)e(n).
\end{equation}
The filtered reference signal vector is given by 
\begin{equation}\label{eq2_19}
    \mathbf{x}'(n)=\hat{s}(n)\ast\mathbf{x}(n)
\end{equation}
where we replace the impulse response $s(n)$ with its estimate $\hat{s}(n)$. Hence, substituting (\ref{eq2_18}) and (\ref{eq2_19}) into (\ref{eq2_17}) yields the update equation of the FxLMS algorithm
\begin{equation}\label{eq2_20}
    \mathbf{w}(n+1) = \mathbf{w}(n) + \mu e(n)\mathbf{x}'(n).
\end{equation}
Given the assumption that the group delay of the secondary path is denoted as $D_\mathrm{s}$, the FxLMS method imposes a constraint on the step size $\mu$.
\begin{equation}\label{eq2_bound}
    0<\mu<\frac{1}{\lambda_\mathrm{max}D_\mathrm{s}}
\end{equation}
where $\lambda_\mathrm{max}$ represents the maximum eigenvalue of the auto-correlation matrix $R_\mathrm{\mathbf{x}'}=\mathbb{E}[\mathbf{x'}^\mathrm{T}(n)\mathbf{x'}(n)]$ of the filtered reference signal. 

As described above, a single-channel feedforward FxLMS algorithm is composed three parts: (1) the control filter, as in (\ref{eq2_15}); (2) the secondary path modeling, as in (\ref{eq2_19}); (3) the coefficients updating, as in (\ref{eq2_20}). Hence, a single channel feedforward FxLMS must complete $3N+1$ multiply operations and $3N-2$ additions in a cycle, assuming that the length of the control filter and secondary path estimate are both $N$. The practical difficulties are resolved through the construction of derivatives of the FxLMS algorithm~\cite{shi2019two,shi2019optimal,shen2023momentum,shi2021comb}, owing to its low computational complexity. 

\subsection{Numerical simulation of the FxLMS algorithm}
The simulation of the FxLMS algorithm is performed using the "dsp.FilteredXLMSFilter" function in the digital signal processing toolbox of MATLAB. The MATLAB code has been included in the Appendix section for the purpose of reference. The spectrum provides confirmation of the algorithm's effectiveness:
\begin{figure}[htbp]
    \centering
    \includegraphics[width=12cm]{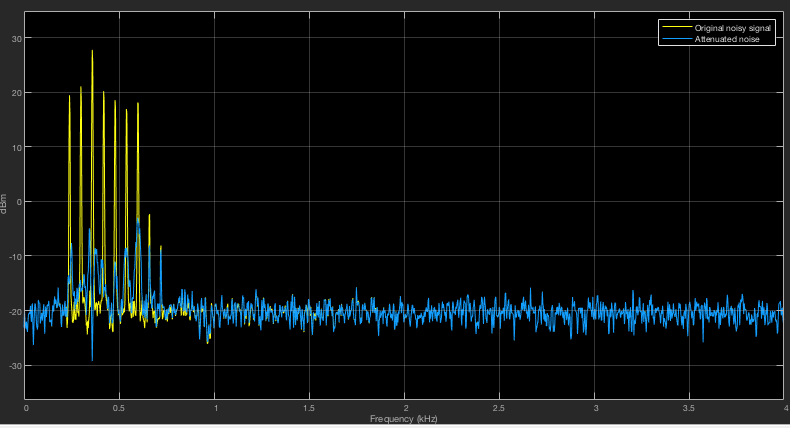}
    \caption{Spectrum of the error signal of active noise control using FXLMS algorithm}
    \label{fig:14}
\end{figure}

There is a notable reduction in the amplitude of the noise when compared to the original signal. The algorithm demonstrates proficiency in estimating the noise source and generating appropriate control signals, as evidenced by the response of both the primary and secondary path.
\begin{figure}[htbp]
    \centering
    \includegraphics[width=11cm]{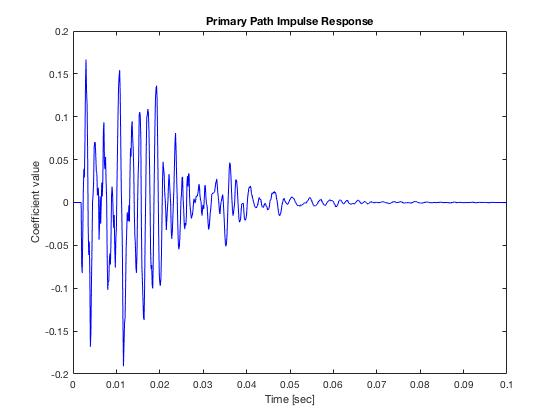}
    \caption{The impulse response of the primary path}
    \label{fig:15}
\end{figure}
\begin{figure}[htbp]
    \centering
    \includegraphics[width=11cm]{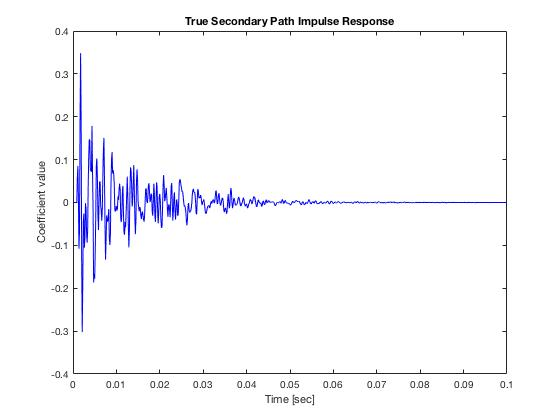}
    \caption{The impulse response of the secondary path}
    \label{fig:16}
\end{figure}
\begin{figure}[htbp]
    \centering
    \includegraphics[width=11cm]{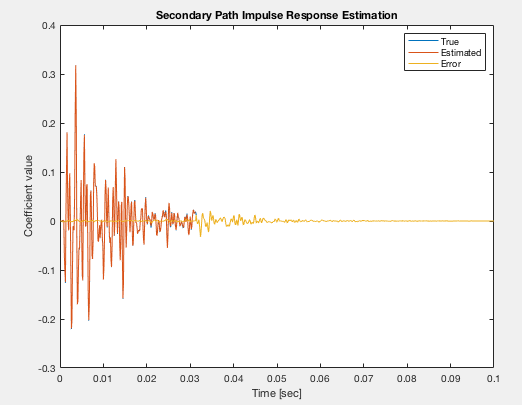}
    \caption{The estimate of the secondary path}
    \label{fig:16}
\end{figure}
\begin{figure}[htbp]
    \centering
    \includegraphics[width=11cm]{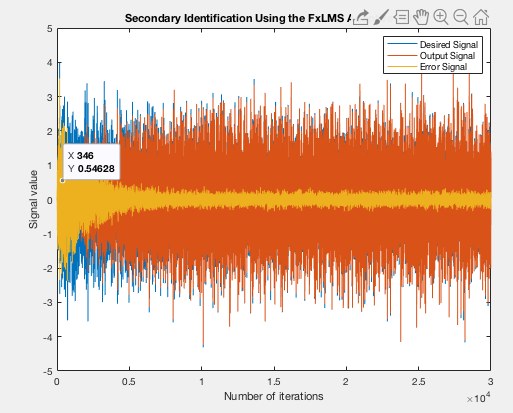}
    \caption{The noise reduction performance of the FxLMS algorithm}
    \label{fig:18}
\end{figure}

Based on the aforementioned data, the estimation of secondary path transfer coefficients demonstrates a notable level of accuracy, which consequently contributes to the effective compression of noise. Nevertheless, the FXLMS simulation exhibits a lengthier processing time compared to the previous LMS simulation, mostly attributed to the intricate nature of noise sources and the utilization of diverse techniques.

\section{Multichannel FxLMS algorithm}
\begin{figure}[htbp]
  \centering
    \includegraphics[scale=1]{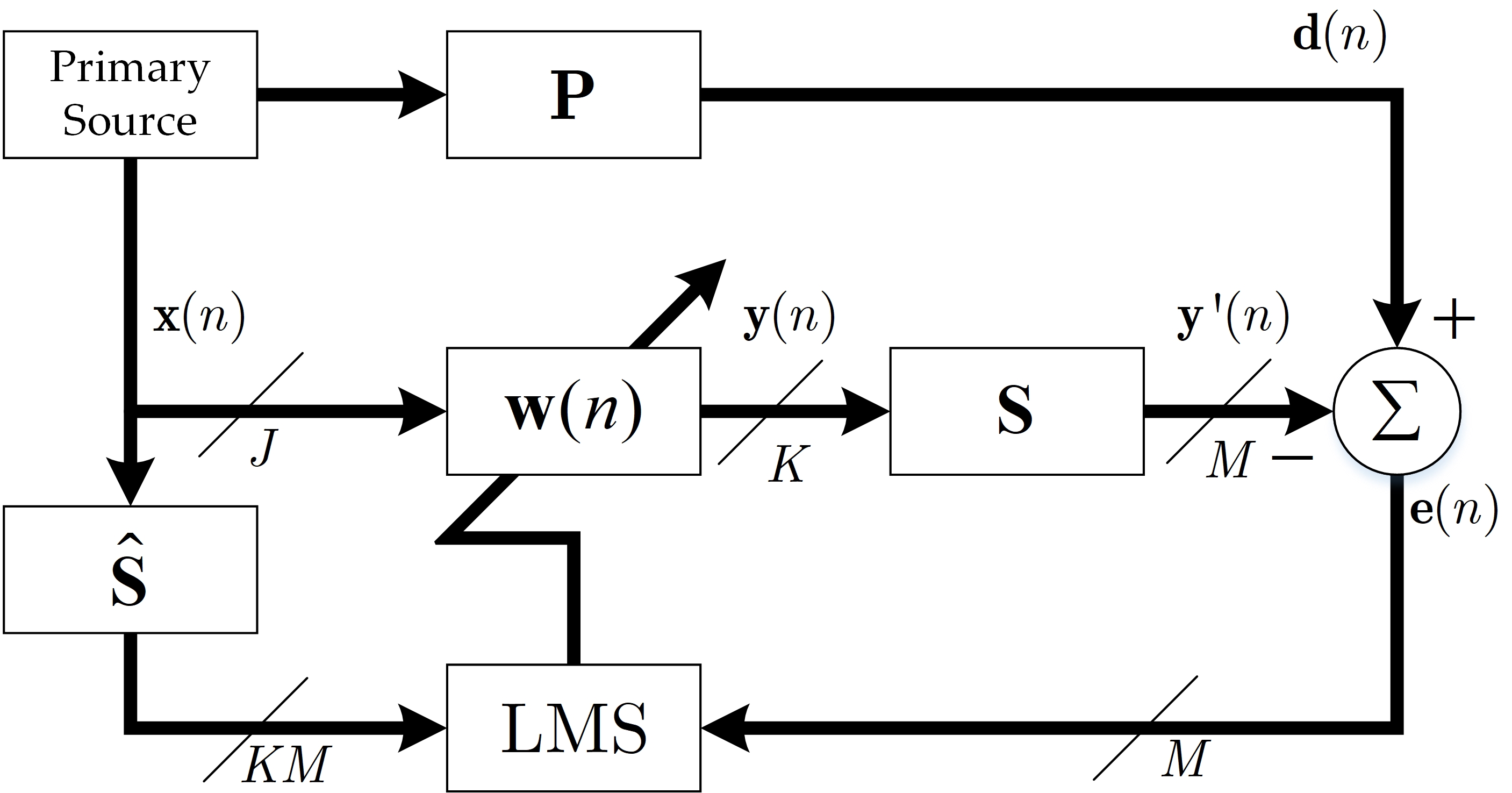}
  \caption{Block diagram of the multichannel FxLMS algorithm.}
  \label{Fig2_4}
\end{figure}
The block diagram of the multichannel FxLMS (MCFxLMS) method is depicted in Figure~\ref{Fig2_4}. This algorithm consists of $J$ reference microphones, $K$ secondary sources, and $M$ error microphones, as described in previous works~\cite{kuo1999active,elliott1987multiple}. The control filter matrix is provided by. 
\begin{equation}\label{eq2_21}
    \mathbf{w}(n)=\begin{bmatrix}
                    \mathbf{w}^\mathrm{T}_\mathrm{11}(n) & \mathbf{w}^\mathrm{T}_\mathrm{12}(n) & \cdots & \mathbf{w}^\mathrm{T}_\mathrm{1J}(n)\\
                    \mathbf{w}^\mathrm{T}_\mathrm{21}(n) & \mathbf{w}^\mathrm{T}_\mathrm{22}(n) & \cdots & \mathbf{w}^\mathrm{T}_\mathrm{2J}(n)\\
                    \vdots                    & \vdots                    & \ddots & \vdots\\
                    \mathbf{w}^\mathrm{T}_\mathrm{K1}(n) & \mathbf{w}^\mathrm{T}_\mathrm{K2}(n) & \cdots & \mathbf{w}^\mathrm{T}_\mathrm{KJ}(n)\\
                  \end{bmatrix}\in\mathbb{R}^\mathrm{K\times JN}
\end{equation}
where $\mathbf{w}_\mathrm{kj}(n)=\left[w_\mathrm{kj,0}(n),w_\mathrm{kj,1}(n),\cdots,w_\mathrm{kj,N-1}(n)\right]^\mathrm{T}\in\mathbb{R}^\mathrm{JN\times 1}$ is the control filter from the $j$th input to the $k$th output, and $N$ denotes the length of the control filter. The reference vector is provided by
\begin{equation}\label{eq2_22}
    \mathbf{x}(n)=\left[\mathbf{x}^\mathrm{T}_1(n),\mathbf{x}^\mathrm{T}_2(n),\cdots,\mathbf{x}^\mathrm{T}_\mathrm{J}(n)\right]^\mathrm{T}\in\mathbb{R}^\mathrm{JN\times 1}
\end{equation}
where $\mathbf{x}_\mathrm{j}=\left[x_\mathrm{j}(n),x_\mathrm{j}(n-1),\cdots,x_\mathrm{j}(n-N+1)\right]^\mathrm{T}\in\mathbb{R}^\mathrm{N\times 1}$. The output of the control filter is given by 
\begin{equation}\label{eq2_23}
    \mathbf{y}(n) = \mathbf{w}(n)\mathbf{x}(n)\in\mathbb{R}^\mathrm{K\times 1}.
\end{equation}
The anti-noise vector at error microphones, as illustrated in Fig.~\ref{Fig2_4}, can be expressed as
\begin{equation}\label{eq2_24}
    \mathbf{y}'(n)=\mathbf{s}\ast\mathbf{y}(n)\in\mathbb{R}^\mathrm{M\times 1}
\end{equation}
where $\mathbf{s}$ represents the impulse response matrix of secondary paths 
\begin{equation}\label{eq2_25}
    \mathbf{s}=\begin{bmatrix}
                    \mathbf{s}_\mathrm{11}(n) & \mathbf{s}_\mathrm{12}(n) & \cdots & \mathbf{s}_\mathrm{1K}(n)\\
                    \mathbf{s}_\mathrm{21}(n) & \mathbf{s}_\mathrm{22}(n) & \cdots & \mathbf{s}_\mathrm{2K}(n)\\
                    \vdots                    & \vdots                    & \ddots & \vdots\\
                    \mathbf{s}_\mathrm{M1}(n) & \mathbf{s}_\mathrm{M2}(n) & \cdots & \mathbf{s}_\mathrm{MK}(n)\\
                  \end{bmatrix}\in\mathbb{R}^\mathrm{M\times K}.
\end{equation}
The error signal vector $\mathbf{e}(n)$ measured by $M$ error microphones can be expressed as
\begin{equation}\label{eq2_26}
    \mathbf{e}(n) = \mathbf{d}(n)-\mathbf{y}'(n)\in\mathbb{R}^\mathrm{M\times 1}
\end{equation}
where $\mathbf{d}(n)=\left[d_1(n), d_2(n), \cdots, d_\mathrm{M}(n)\right]^\mathrm{T}\in\mathbb{R}^\mathrm{M\times 1}$ stands for the disturbance vector, with $d_m(n)$ denotes the disturbance at the $m$th ($m=1,2,\cdots,M$) error microphone. The cost function of the adaptive filter is defined as 
\begin{equation}\label{eq2_27}
    J(n)=\sum^\mathrm{M}_\mathrm{m=1}e^2_\mathrm{m}(n).
\end{equation}
Hence, the gradient of the cost function is
\begin{equation}
\begin{split}
    \nabla &J(n)=\\&\begin{bmatrix}
                    \sum^\mathrm{M}_\mathrm{m=1}e_\mathrm{m}\left[s_\mathrm{m1}\ast\mathbf{x}_1(n)\right] & \sum^\mathrm{M}_\mathrm{m=1}e_\mathrm{m}\left[s_\mathrm{m1}\ast\mathbf{x}_2(n)\right] & \cdots & \sum^\mathrm{M}_\mathrm{m=1}e_\mathrm{m}\left[s_\mathrm{m1}\ast\mathbf{x}_\mathrm{J}(n)\right]\\
                    \sum^\mathrm{M}_\mathrm{m=1}e_\mathrm{m}\left[s_\mathrm{m2}\ast\mathbf{x}_1(n)\right] & \sum^\mathrm{M}_\mathrm{m=1}e_\mathrm{m}\left[s_\mathrm{m2}\ast\mathbf{x}_2(n)\right] & \cdots & \sum^\mathrm{M}_\mathrm{m=1}e_\mathrm{m}\left[s_\mathrm{m2}\ast\mathbf{x}_\mathrm{J}(n)\right]\\
                    \vdots                    & \vdots                    & \ddots & \vdots\\
                    \sum^\mathrm{M}_\mathrm{m=1}e_\mathrm{m}\left[s_\mathrm{mK}\ast\mathbf{x}_1(n)\right] & \sum^\mathrm{M}_\mathrm{m=1}e_\mathrm{m}\left[s_\mathrm{mK}\ast\mathbf{x}_2(n)\right] & \cdots & \sum^\mathrm{M}_\mathrm{m=1}e_\mathrm{m}\left[s_\mathrm{mK}\ast\mathbf{x}_\mathrm{J}(n)\right]\\
                  \end{bmatrix}\notag
\end{split}
\end{equation}
where we replace the impulse response $s_\mathrm{mk}(n)$ with its estimate $\hat{s}_\mathrm{mk}(n)$ to give
\begin{equation}\label{eq2_28}
    x^\prime_\mathrm{jkm}(n)=\hat{s}_\mathrm{mk}(n)\ast x_\mathrm{j}(n)
\end{equation}
The control filter weight vector $\mathbf{w}_\mathrm{kj}(n)$ is updated in the direction of the negative gradient with step size $\mu$ based on the minimization of the estimated cost function of (\ref{eq2_27}):
\begin{equation}\label{eq2_29}
    \mathbf{w}_\mathrm{kj}(n+1) = \mathbf{w}_\mathrm{kj}(n)+\mu\sum^\mathrm{M}_\mathrm{m=1}e_m(n)\mathbf{x}'_\mathrm{jkm}(n)
\end{equation}
where the filtered reference vector is
\begin{equation}
   \mathbf{x}'_\mathbf{jkm}(n)=\left[x'_\mathrm{jkm}(n),x'_\mathrm{jkm}(n-1),\cdots,x'_\mathrm{jkm}(n-N+1)\right]^\mathrm{T}.\notag 
\end{equation}
From the above analysis, the number of multiplications and additions in MCFxLMS algorithm for one cycle is given by
\begin{equation}\label{eq2_30}
    N_\mathrm{0}=JKM\left(N+L_\mathrm{s}\right)+JKN
\end{equation}
where $L_\mathrm{s}$ denotes the length of the secondary path estimate. If $J=K=M=L$ and $L_\mathrm{s}=N$, the number of multiplications and additions becomes
\begin{equation}\label{eq2_31}
    N_\mathrm{0}=L^2N\left(2L+1\right).
\end{equation}
As shown from Equation (\ref{eq2_31}), it is apparent that the computational burden experiences exponential growth as the number of MCFxLMS channels rises. In recent times, several computationally efficient multichannel adaptive algorithms~\cite{shi2019practical,shi2021block,shi2020active,shi2023multichannel,shi2023computation} have been devised with the aim of reducing the computing burden associated with the multichannel ANC adaptive algorithm. 

\section{Modified FxLMS algorithm}
Based on the concept of FxLMS presented in the preceding section, the secondary route modeling process involves the incorporation of an additional adaptive filter into the system. This inclusion, however, leads to the introduction of delay and increased computational complexity. Consequently, the modified Filtered-X Least Mean Squares (FxLMS) method~\cite{akhtar2009improving,lai2023mov} incorporates two additional filters in the secondary path, as opposed to the utilization of adaptive filters. The intermediate step involves the computation of the disturbance segment within the error signal, resulting from the control source, represented as $d(n)$. Subsequently, the adjusted error signal $e_m(n)$ is provided to the LMS block for the purpose of updating the control filter coefficients~\cite{akhtar2004modified}. The formula for determining the disturbance signal, denoted as $d(n)$, is provided as follows:
\begin{equation}
    {d}(n)=e(n)-\sum^{J-1}_{j=0}\hat{s}_j y(n-j)
\end{equation}
whereas $\hat{S}$ stands for the transfer function, its $z$-transform is $S(z)$, and $j$ is its order number, which is illustrated before. The error signal in the FxLMS algorithm can also be expressed in the form of:
\begin{equation}
    e(n)=d(n)+\sum^{J-1}_{j=0}\sum^{I-1}_{i=0}s_j\cdot w_i(n-j)x(n-i-j)
\end{equation}
Given the assumption of delayed adaptation, it is possible to reframe the error signal as follows:
\begin{equation}
    e_m(n)=d(n)+\sum^{J-1}_{j=0}\sum^{I-1}_{i=0}s_j\cdot w_i(n)x(n-i-j)
\end{equation}
The formula demonstrates that the modified error signal is dependent on the present filter coefficients. Lastly, similar to the FxLMS method, the update function for the filter coefficients is expressed as:
\begin{equation}
    \mathbf{w}(n+1)=\mathbf{w}(n)-\mu e_m(n)\mathbf{x}_f(n)
\end{equation}
where $\mathbf{x}_f(n)$ represents the filtered reference signal. 

Figure 19 presents a schematic diagram that illustrates the function blocks of the modified FxLMS algorithm. The physical simulation is conducted using the Sigma Studio software and will be further elaborated upon in subsequent discussions.
\begin{figure}[htbp]
  \centering
    \includegraphics[width=14cm]{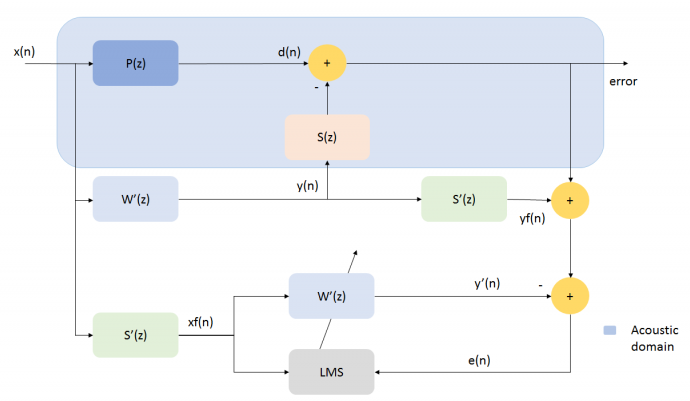}
  \caption{Block diagram of the modified FxLMS algorithm.}
  \label{Fig2_5}
\end{figure}
\section{Latest adaptive noise control algorithms}
In order to address the practical issues encountered in ANC applications and mitigate certain limitations associated with conventional adaptive methodologies, numerous innovative techniques and algorithms have been recently presented, such as some deep-learning-based noise control algorithms~\cite{luo2022implementation,shi2022selective,luo2022hybrid,luo2023performance,shi2023transferable,luo2023deep}, to accelerate the convergence~\cite{shi2021fast}, increase the stability of the ANC system~\cite{gan2023practical}, alleviate the computational complexity~\cite{wen2020using,shi2018novel,shi2020feedforward}, enlarge the noise cancellation zone~\cite{shi2020multichannel,shi2019analysis,lai2023real}, enhance the perception effect of the users, and so on~\cite{lai2023real,lai2023robust}.
\lhead{Introduction to Development Tools}
\chapter{3. Introduction to development tools}
\section{Sigma Studio}
SigmaStudio is a software application developed by Analog Devices with the purpose of facilitating the programming of SigmaDSP audio processors through a user-friendly graphical interface. The introduction of Sigma Studio has significantly enhanced the efficiency of virtual Pseudo-instrument development and minimized the programming demands for developers. Additionally, the organization of Sigma Studio projects is orderly and uncomplicated, allowing for direct modifications at each function block. In summary, it is user-friendly and easily graspable. The project developed is directly linked to a physical test bench, exhibiting exceptional data transmission speed and a lack of defects or errors. Sigma Studio is a software platform that mostly centers around audio processing. It offers a wide range of embedded algorithm function blocks, which have been specifically built to cater to the requirements of various integrated circuits (ICs). Additionally, Sigma Studio offers a comprehensive solution for addressing many challenges, including audio channel segmentation, audio data input and output, and the exportation of learned filters and experimental outcomes. A standard Sigma Studio project is comprised of Nodes, Terminals, and Wires. The execution elements within the context of this discussion are commonly referred to as nodes or functions. These nodes or functions have resemblance to the code and functions found in textual programming languages. Terminals can be conceptualized as the interfaces via which nodes receive input and deliver output. The transmission of data between nodes or terminals occurs via wired connections. In practical applications, it is imperative to give careful consideration to the many features of data, such as the sampling rate, in order to ensure the accurate and seamless transmission of information. Finally, a variety of examples of digital signal processing (DSP) methods and tools are presented. These include the LMS (Least Mean Squares) and NLMS (Normalized Least Mean Squares) algorithms, operations such as muting and amplifying sound tracks, as well as signal mixing and filtering. Furthermore, this technology empowers users to create and implement their own algorithms using Field-Programmable Gate Array (FPGA) circuits. The program has many integrated digital signal processing modules. Additionally, the ADAU1452 chip possesses certain functionalities. It should be noted that the featured applications may vary depending on the selection of a different chip. In this scenario, the majority of fundamental active noise control applications can be directly implemented and simulated.
\begin{figure}
    \centering
    \includegraphics[width=6.5cm]{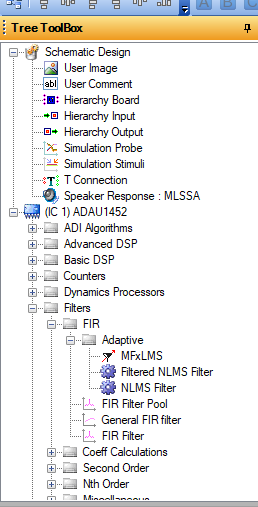}
    \caption{Display of Sigma Studio function blocks}
    \label{fig:sigmastudio}
\end{figure}

\section{ADAU1452 chip}

\begin{figure}
    \centering
    \includegraphics[width=15cm]{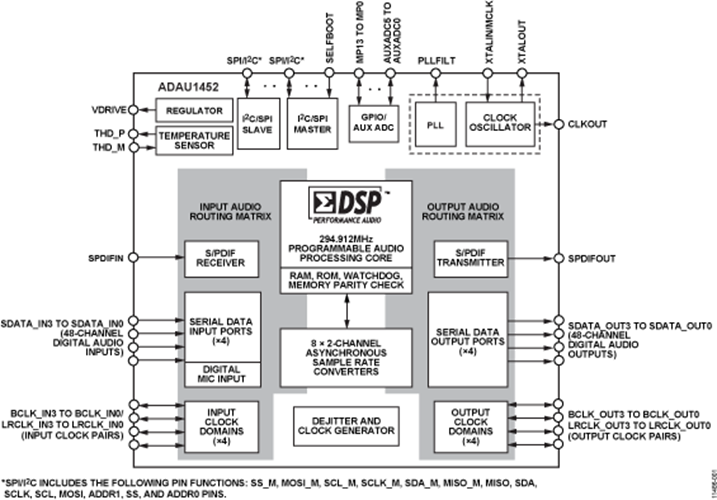}
    \caption{The schematic of the ADAU1452 chip}
    \label{fig:ADAU1452}
\end{figure}

The ADAU1452 is a programmable signal processing chip manufactured by Analog Devices. It is specifically designed for digital signal processing and is well-suited for automation applications~\cite{ADAU1452,ADAU1452_EVM}. Typical methods encompass the utilization of digital signal amplifiers and automobile audio processing, alongside the implementation of active noise control techniques. The processing speed of the latest integrated circuit (IC) designed by Analog Devices has been significantly improved. This is achieved by employing a sample by sample, block by block approach, which is based on the principle of divide and conquer. The relationship between the IC board and Sigma studio has been enhanced, enabling a wider range of applications to be executed using graphical coding methods that offer improved precision and ease. 

Various functions like as digital to analog conversion (DAC), analog to digital conversion (ADC), amplification, and the associated control circuits, as well as signal transmission via pulse density modulation or pulse code modulation, can be readily accomplished with minimal additional effort.  Subsequently, the chip's parameters are enumerated as follows. The system is equipped with a 32-bit, 1.2V central processing unit (CPU) specifically designed for signal processing tasks. The CPU operates at a working frequency of 294.912MHz and a sampling frequency of 48kHz. Additionally, it has the capability to support 4-port and 48-channel inputs and outputs, offering a maximum sampling rate of 192kHz. Furthermore, the device provides support for both 3D sound input and output. It is also equipped with eight ASRCs (Asynchronous Sample Rate Converters) and has the capability to achieve a dynamic range of 139 decibels (dB) for its signal-to-noise ratio (DNR). Moreover, it exhibits a higher level of energy efficiency in comparison to its predecessors.
\lhead{Modified-FXLMS Experiment}
\chapter{4. Modified FxLMS Experiments}
\section{Modified FxLMS algorithm physical test in duct (closed environment)}
Due to its high computational efficiency and excellent convergence behavior, the modified FxLMS algorithm has been employed in this part to mitigate the noise present in the duct.  

\subsection{Experimental system setup}
\begin{figure}[htbp]
  \centering
    \includegraphics[width=14cm]{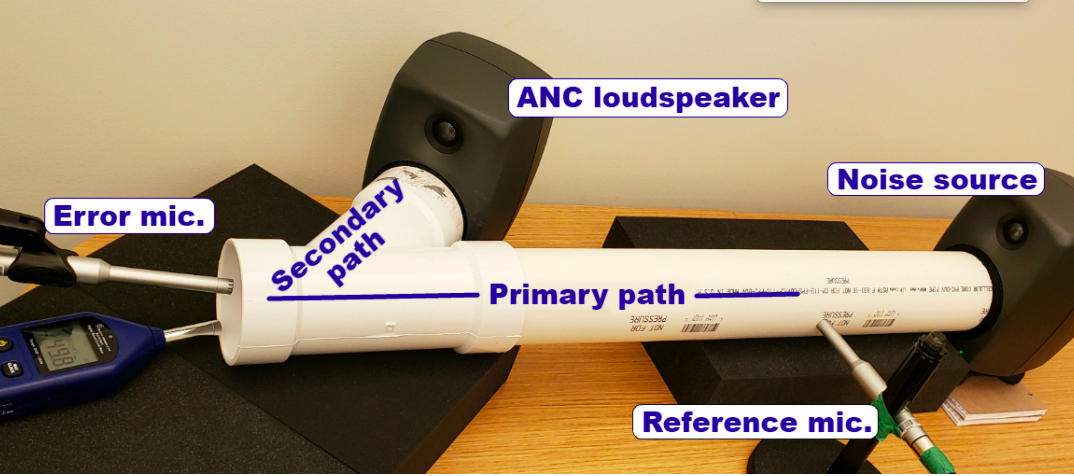}
  \caption{The experimental configuration for the test bench.}
  \label{Fig21}
\end{figure}

\begin{figure}[htbp]
  \centering
    \includegraphics[width=14cm]{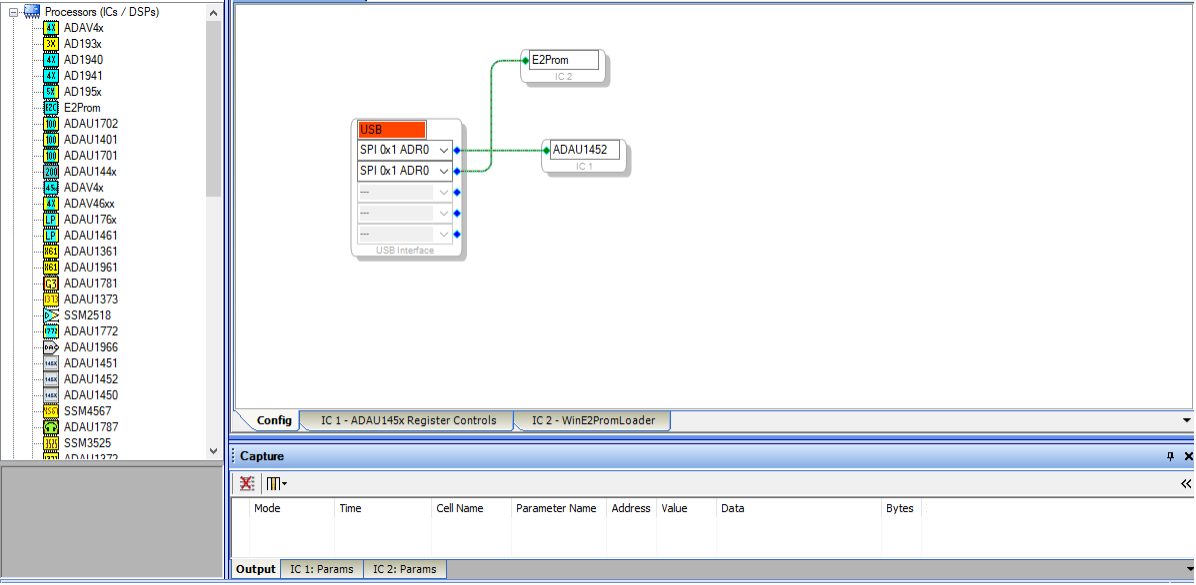}
  \caption{Sigma Studio hardware connection.}
  \label{Fig22}
\end{figure}

\begin{figure}[htbp]
  \centering
    \includegraphics[width=13cm]{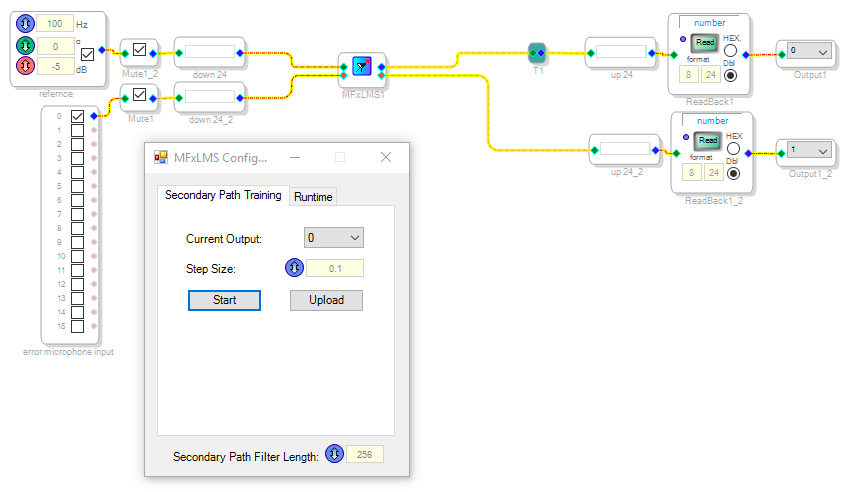}
  \caption{Sigma Studio software set up.}
  \label{Fig23}
\end{figure}

The construction of the test bench is based on the guidelines provided by the official MATLAB website, as shown in Figure~\ref{Fig21}. The subsequent paragraph provides an introduction to the components utilized in the tests and their respective configurations. The elongated cylindrical structure represents the principal conduit, denoting the acoustic component of the adapted FXLMS function block. The system comprises of a noise generator, a reference microphone, and an error microphone. The secondary path refers to a specific branch of the pipe that is equipped with a control source located at its opening. In the conducted experiment, the designated position of the noise source is indicated as 0 cm. The reference microphone is situated at a distance of 8 cm, while the anti-noise source, serving as the initiation point of the secondary path, is positioned at 73 cm. Lastly, the error microphone is put at a distance of 78 cm, representing the termination point of the pipe. The placement of a cotton material behind the reference microphone serves the purpose of preventing the unwanted transmission or interference of the anti-noise signal, hence ensuring the integrity and precision of the reference signal. The software and hardware configuration in SigmaStudio are depicted in Figure~\ref{Fig22} and Figure~\ref{Fig23}, respectively.

The hardware configuration involves the connection of an EEProm and an ADAU control unit, facilitating the implementation of real-time active noise reduction. The purpose of EEPROM is to serve as a temporary storage medium for the signal acquired by microphones. In contrast, ADAU1452 is responsible for doing the necessary computations for the updated FxLMS algorithm and supplying the control source with the resulting output signal. 

The aforementioned software configuration represents a single-channel active noise control (ANC) system. The ADAU1452 chip utilizes two input channels to transmit a reference signal and an error signal to a modified FXLMS function block. This function block incorporates the noise acquired by two microphones.

The modified FXLMS function block is equipped with two distinct operational modes, namely the training mode and the runtime mode. The purpose of the training mode is to calculate the secondary path coefficient, which is then utilized to mitigate the impact of secondary path disturbances. The purpose of the runtime mode is to execute the modified FXLMS algorithm in real-time signal processing. During the training mode, the input for the error signal and the input for the reference signal are both deactivated. Consequently, the signal processing unit exclusively gets the stochastic noise generated at the control source, as well as the ensuing noise captured by the error microphone. The secondary path coefficient can be defined as the ratio between the two signals.

A mute block is implemented at channel 2, specifically at input 16, in order to silence the reference microphone when required. This action is taken, for example, when the training mode is activated or when troubleshooting is necessary. The speaker is connected to channel 1, which represents the anti-noise signal, for the purpose of generating outputs. The additional output is not relevant to the sigma studio project and is deactivated in subsequent experiments.
In order to ascertain the correct connection of the signal and microphones to their respective ports, a preliminary test is undertaken.
\begin{figure}[htbp]
  \centering
    \includegraphics[width=13cm]{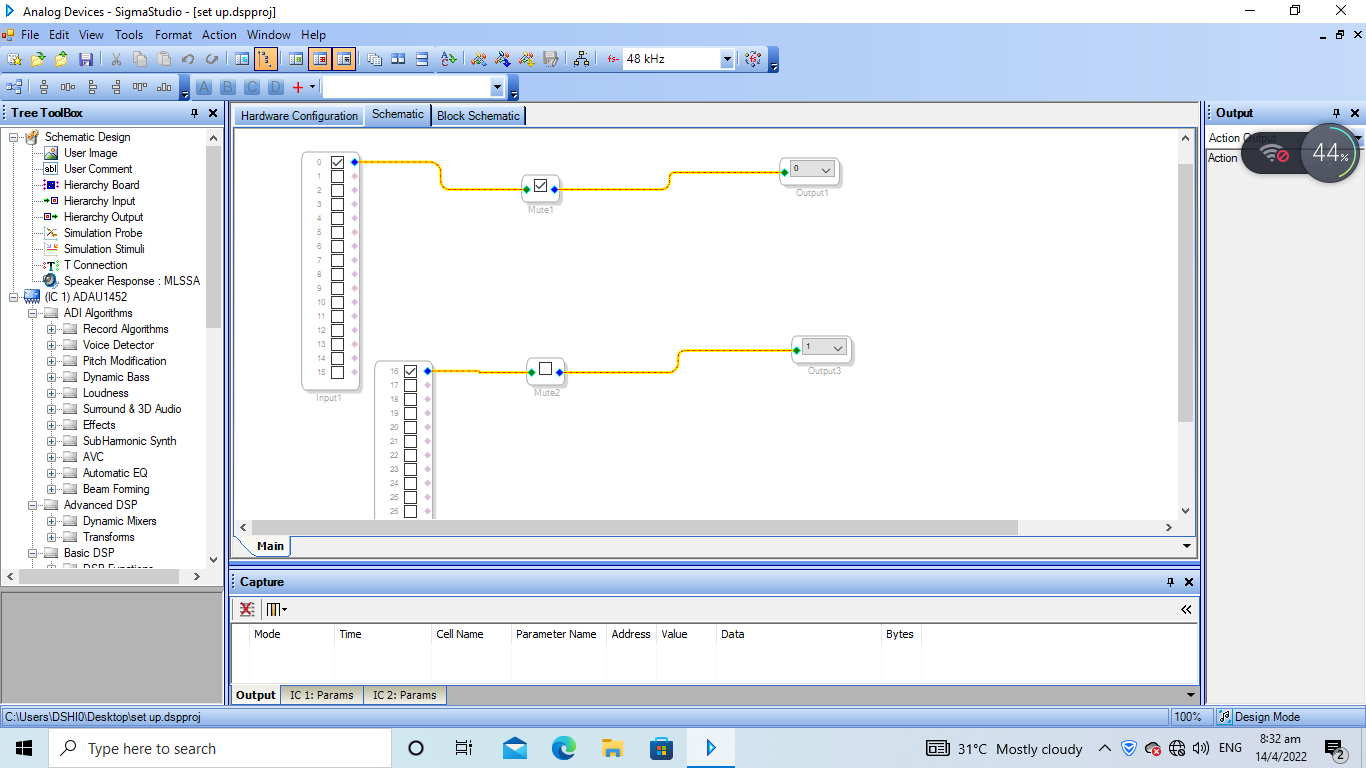}
  \caption{Channel connection test.}
  \label{Fig24}
\end{figure}

Loudspeakers and microphones are typically linked to respective output and input pins. In the context of loudspeaker testing, the computer is often linked to the input pin. The accuracy of the connection is determined by the successful production of the intended sound. Microphones are typically linked to computers in order to capture audio signals produced by the computer. When a microphone is properly plugged in, it effectively eliminates any extraneous noise. The validation of the connection between the loudspeakers and the ADAU board has recently been completed. The functionality of this connection may be determined by assessing if the loudspeaker is producing the intended sound. In order to establish proper connections between amplifiers and speakers, it is imperative to exercise care and meticulousness, as both devices possess distinct positive and negative ports that necessitate appropriate interconnections.

\subsection{Experiment process}
The objective of the studies is to determine the optimal settings for the function block, as well as identify the most effective arrangement for the test bench. 

The determination of the filter length, step size, and sampling frequency is required. The determination of the final outcome will be contingent upon the findings obtained from the conducted experiment. The initial phase of the experiment involves training the adaptive filter. Prior to commencing the training process, both input signals are silenced to facilitate the acquisition of the secondary path filter S(z). Once the trained filter has been saved, it can be utilized for testing purposes.

During the training process of the secondary path, the coefficients of the secondary path are trained using three different filter lengths: 32, 64, and 128. The default step size for training is set at 0.9, and the duration of the training session is 30 seconds. The outcomes are stored in a text file for subsequent reference in the run-time mode.

The control variable method is employed during the testing process. The placement of microphones and speakers is predetermined and unalterable. The speaker in the pipe serves as the noise source, emitting a range of noises produced by Audacity. Given that the sampling rate in our specific scenario is 48kHz, and the signal is down-sampled by a factor of 12 before entering the modified FxLMS block, it can be concluded that the effectiveness of the active noise control (ANC) system is limited when it comes to attenuating high frequency noise. However, it was seen that the active noise control (ANC) system was not efficient in attenuating low-frequency noises during the experiment. Consequently, the noise frequencies selected for testing were 250 Hz, 3000 Hz, 400 Hz, 500 Hz, and 600 Hz, as these were the frequencies that could be effectively suppressed by the ANC system. The quantification of noise reduction is determined by comparing the readings obtained from the spectrometer positioned at the terminus of the pipe both prior to and subsequent to the implementation of active noise control. The tables below display the test results obtained under three different situations.

\begin{table}[htbp]
\caption{Filter length: $32$, step size: $10^{-5}$}
\begin{tabular}{|l|l|l|l|}
\hline
\multicolumn{1}{|c|}{\textbf{Frequency}} & \multicolumn{1}{c|}{\textbf{Noise level (dBA) before ANC}} & \multicolumn{1}{c|}{\textbf{Noise level (dBA)  after ANC}} & \multicolumn{1}{c|}{\textbf{NR}} \\ \hline
250hz                                    & 75                                                               & Diverge                                                          & Fail                                          \\ \hline
300hz                                    & 75                                                               & 65                                                               & 10                                            \\ \hline
400hz                                    & 78                                                               & Diverge                                                          & Fail                                          \\ \hline
500hz                                    & 78                                                               & Drop to 72,then Diverge.                                         & Fail                                          \\ \hline
600hz                                    & 76                                                               & 60                                                               & 16                                            \\ \hline
\end{tabular}
\end{table}

\begin{table}[htbp]
\caption{Filter length: $32$, step size: $10^{-6}$}
\begin{tabular}{|c|c|c|c|}
\hline
\textbf{Frequency} & \textbf{Noise level (dBA) before ANC} & \textbf{Noise level (dBA)  ANC} & \textbf{NR} \\ \hline
250hz              & 75                                    & Diverge                         & Fail        \\ \hline
300hz              & 75                                    & 53                              & 22          \\ \hline
400hz              & 77                                    & Diverge                         & Fail        \\ \hline
500hz              & 78                                    & 57                              & 21          \\ \hline
600hz              & 76                                    & 60                              & 16          \\ \hline
\end{tabular}
\end{table}

\begin{table}[htbp]
\caption{Filter length: $64$, step size: $10^{-5}$}
\begin{tabular}{|c|c|c|c|}
\hline
\textbf{Frequency} & \textbf{Noise level (dBA) before ANC} & \textbf{Noise level (dBA)  ANC} & \textbf{NR} \\ \hline
250hz              & 75                                    & 68                              & 7           \\ \hline
300hz              & 77                                    & Diverge                         & Fail        \\ \hline
400hz              & 83                                    & 73                              & 10          \\ \hline
500hz              & 82                                    & Diverge                         & Fail        \\ \hline
600hz              & 84                                    & Diverge                         & Fail        \\ \hline
\end{tabular}
\end{table}

\begin{table}[htbp]
\caption{Filter length: $64$, step size: $10^{-6}$}
\begin{tabular}{|c|c|c|c|}
\hline
\textbf{Frequency} & \textbf{Noise level (dBA) before ANC} & \textbf{Noise level (dBA)  ANC} & \textbf{NR} \\ \hline
250hz              & 75                                    & 65                              & 10          \\ \hline
300hz              & 77                                    & Diverge, but very slow.         & Fail        \\ \hline
400hz              & 83                                    & 66                              & 17          \\ \hline
500hz              & 82                                    & Diverge                         & Fail        \\ \hline
600hz              & 84                                    & Diverge                         & Fail        \\ \hline
\end{tabular}
\end{table}

\begin{table}[htbp]
\caption{Filter length: $128$, step size: $10^{-5}$}
\begin{tabular}{|c|c|c|c|}
\hline
\textbf{Frequency} & \textbf{Noise level (dBA) before ANC} & \textbf{Noise level (dBA)  ANC} & \textbf{NR} \\ \hline
250hz              & 81                                    & 74                              & 7           \\ \hline
300hz              & 76                                    & 71                              & 5           \\ \hline
400hz              & 83                                    & 80, but not stable              & 3           \\ \hline
500hz              & 81                                    & 79                              & 2           \\ \hline
600hz              & 83                                    & Diverge                         & Fail        \\ \hline
\end{tabular}
\end{table}

\begin{table}[htbp]
\caption{Filter length: $128$, step size: $10^{-6}$}
\begin{tabular}{|c|c|c|c|}
\hline
\textbf{Frequency} & \textbf{Noise level (dBA) before ANC} & \textbf{Noise level (dBA)  ANC} & \textbf{NR} \\ \hline
250hz              & 81                                    & Diverge, but very slow          & Fail        \\ \hline
300hz              & 76                                    & 56                              & 20          \\ \hline
400hz              & 83                                    & 65                              & 18          \\ \hline
500hz              & 81                                    & 64                              & 17          \\ \hline
600hz              & 83                                    & Diverge, but very slow          & Fail        \\ \hline
\end{tabular}
\end{table}

\begin{figure}[htbp]
  \centering
    \includegraphics[width=11cm]{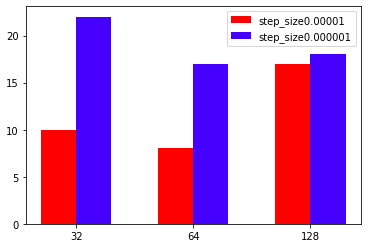}
  \caption{Noise performance at $300$ Hz tonal noise.}
  \label{Fig25}
\end{figure}

\begin{figure}[htbp]
  \centering
    \includegraphics[width=11cm]{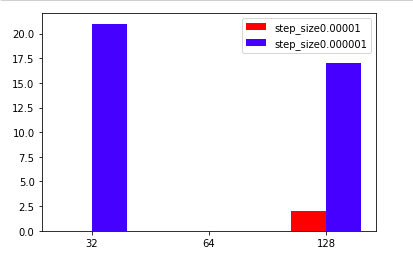}
  \caption{Noise performance at $500$ Hz tonal noise.}
  \label{Fig26}
\end{figure}

The efficacy of each sort of parameter is demonstrated by constructing histograms that display the outcomes of the filters on lowering noises of frequencies $300$ Hz and $500$ Hz, as depicted in Figure~\ref{Fig25} and Figure~\ref{Fig26}.

It is evident that increasing the length of the filter leads to improved stability in performance. This is due to the occurrence of random divergence at certain frequencies when using shorter filter lengths. A filter length of $128$ has been chosen, together with a step size of $10^{-6}$, in order to achieve improved noise reduction.

In conclusion, the system is subjected to testing in the presence of broadband noise. This noise is generated by two sources: the first source is white noise produced by Audacity, with a frequency range of $350$ Hz to $450$ Hz. The second source consists of a combination of tonal noise at $300$ Hz and tonal noise at $400$ Hz.

The initial white noise exhibits a loudness range of 65 to 67 dB. However, following the implementation of the modified FXLMS unit, the loudness is effectively decreased to a range of 54 to 56 dB. Additionally, the combination of tonal noises experiences a reduction from 89 dB to 70 dB. Overall, the noise control demonstrates reliable performance. In general, it can be inferred that the current active noise control (ANC) unit demonstrates efficacy across the frequency range of $300$ Hz to approximately $500$ Hz. 

Several potential limits and downsides have been identified and are presented in the following list:
\begin{enumerate}
    \item The noise source fails to generate noise of superior quality, hence impeding our ability to accurately assess the efficacy of the ANC system.
    \item The magnitude of measurement inaccuracy is significant due to the manual handling of the sound meter, which inevitably leads to variations in its position during several experimental trials. It is plausible that the optimal arrangement of components has yet to be achieved.
\end{enumerate}

\section{Modified FxLMS algorithm test in open space}
\subsection{Physical Set-up}
\begin{figure}[htbp]
  \centering
    \includegraphics[width=13cm]{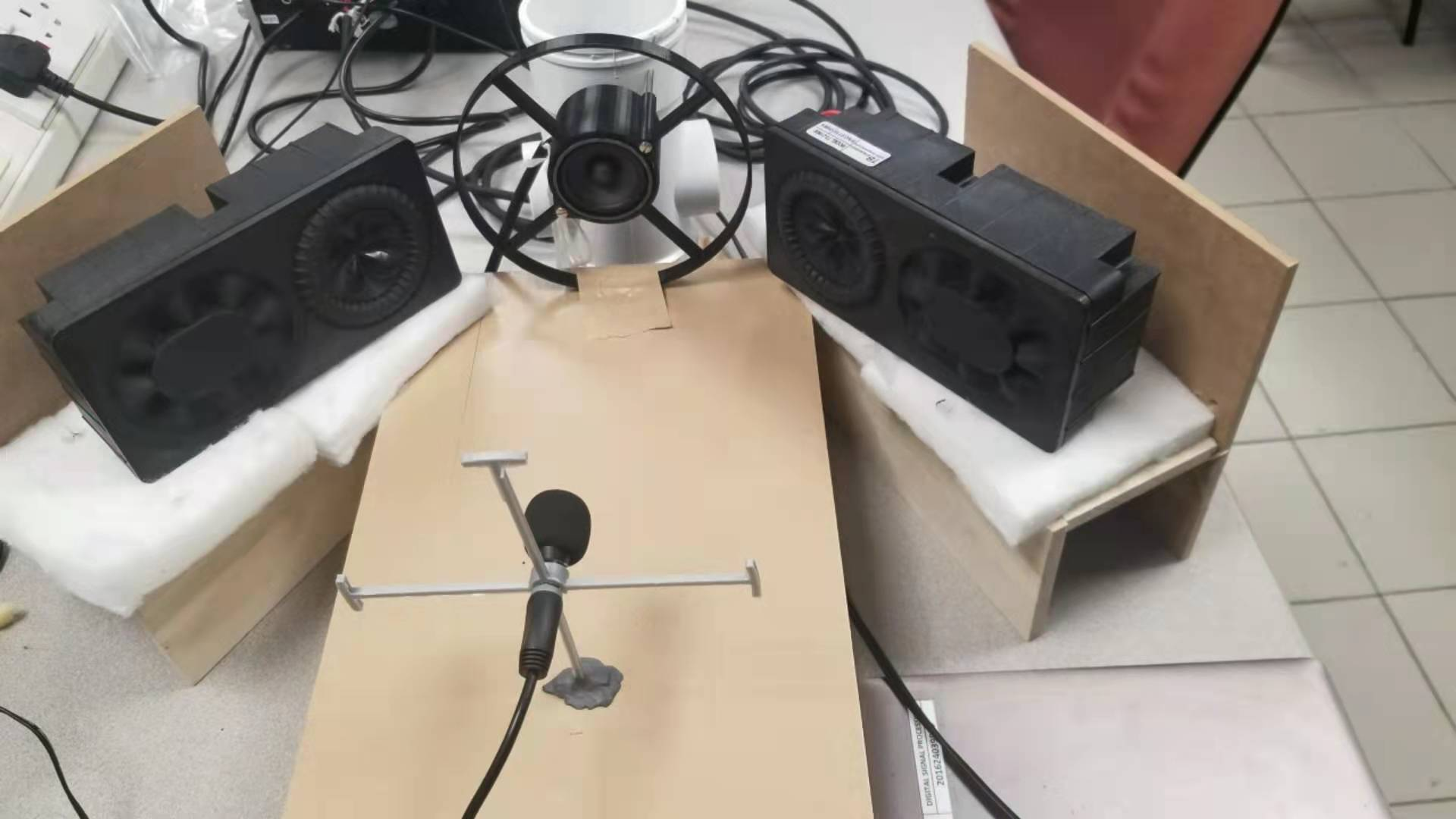}
  \caption{Test bench for open space modified FxLMS ANC}
  \label{Fig27}
\end{figure}

A wooden structure is constructed to accommodate the components, as the outcome of solely providing a single anti-noise source is highly unfavorable. In an attempt to provide a more comprehensive analysis, efforts were made to position the anti-noise speaker at various orientations surrounding the platform. However, no discernible reduction in noise levels was seen. Therefore, the decision has been made to employ a pair of anti-noise sources positioned symmetrically to the noise source, as the outputs of both speakers are identical. In other terms, the objective is to develop a dual-channel active noise control system. Furthermore, as a result of the vibrations generated by the anti-noise speakers, an additional auditory output is produced when the speakers come into contact with the test bench. To mitigate this issue, two wooden stands have been constructed as a precautionary measure. The software logic remains consistent with earlier instances, although the noise source introduces recorded coffee machine noise. The photograph presented in Figure~\ref{Fig27} illustrates the posture that is reasonably ideal. 

In order to provide additional information, it should be noted that the error microphone is positioned at a distance of approximately 40 cm from the control sources. In the event that a reference microphone is present, it is positioned directly in front of the noise source. Additionally, a technique is introduced wherein the reference signal is directly connected to the computer, which serves as the noise generator. This approach aims to enhance the accuracy of the reference signal.

\subsection{Software set-up}
As demonstrated previously, the quantity of control sources has been augmented to two, necessitating corresponding adjustments within the Sigma studio software. In order to achieve this, the outputs of two anti-noise speakers have been selected to be identical. Consequently, a sound spliter has been connected to the output terminal of the modified FxLMS block. To mitigate measurement inaccuracies and enhance computational efficiency, the noise signal is directly linked to the board via USB, originating from the computer, as opposed to using a reference microphone. The comprehensive project is displayed in Figure~\ref{Fig28}, Figure~\ref{Fig29}, and Figure~\ref{Fig30} below.  
\begin{figure}[htbp]
  \centering
    \includegraphics[width=13cm]{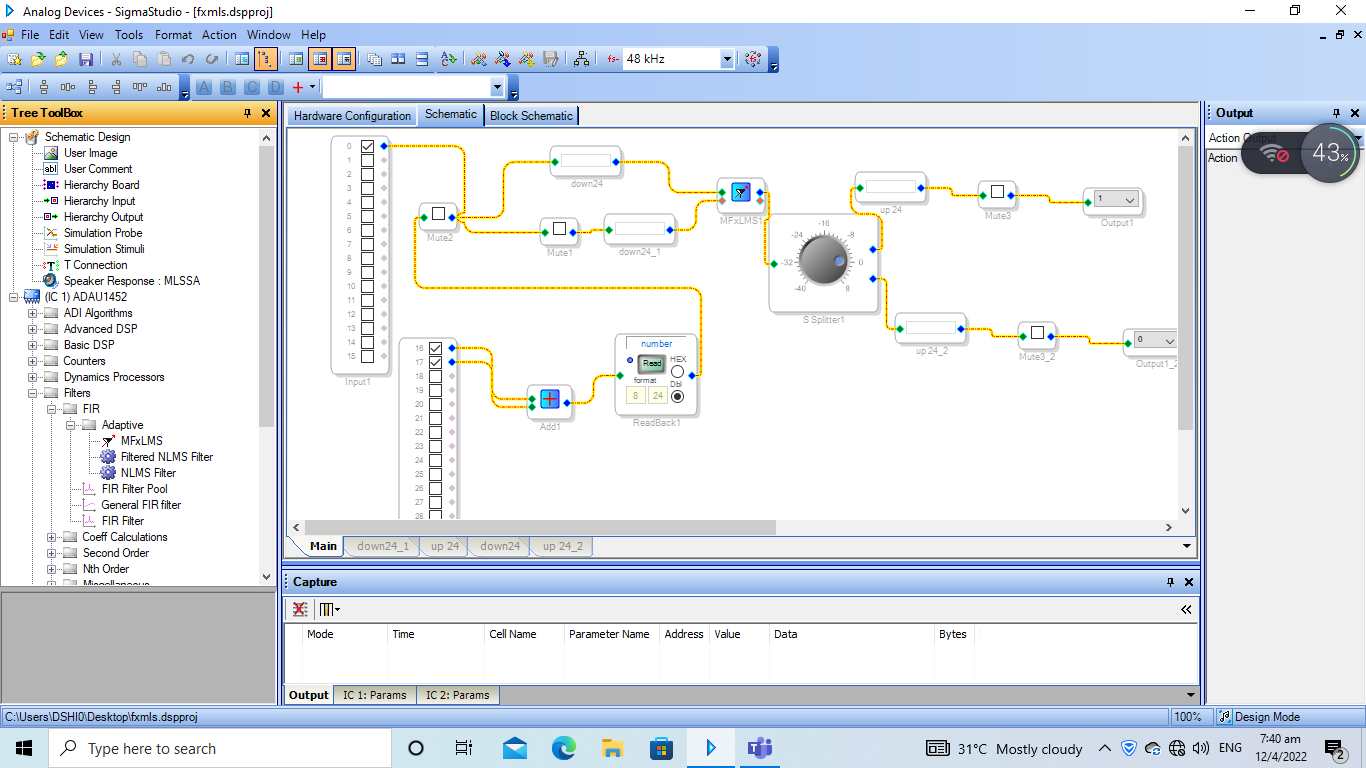}
  \caption{Complete software set up for two control sources modified FxLMS}
  \label{Fig28}
\end{figure}

The supported function blocks regulate the sampling rate of the incoming data to the modified FxLMS block and the outgoing data to the control sources. The $down-24$ downsampling function and $up-24$ upsampling function are the subject of inquiry.
\begin{figure}[htbp]
  \centering
    \includegraphics[width=13cm]{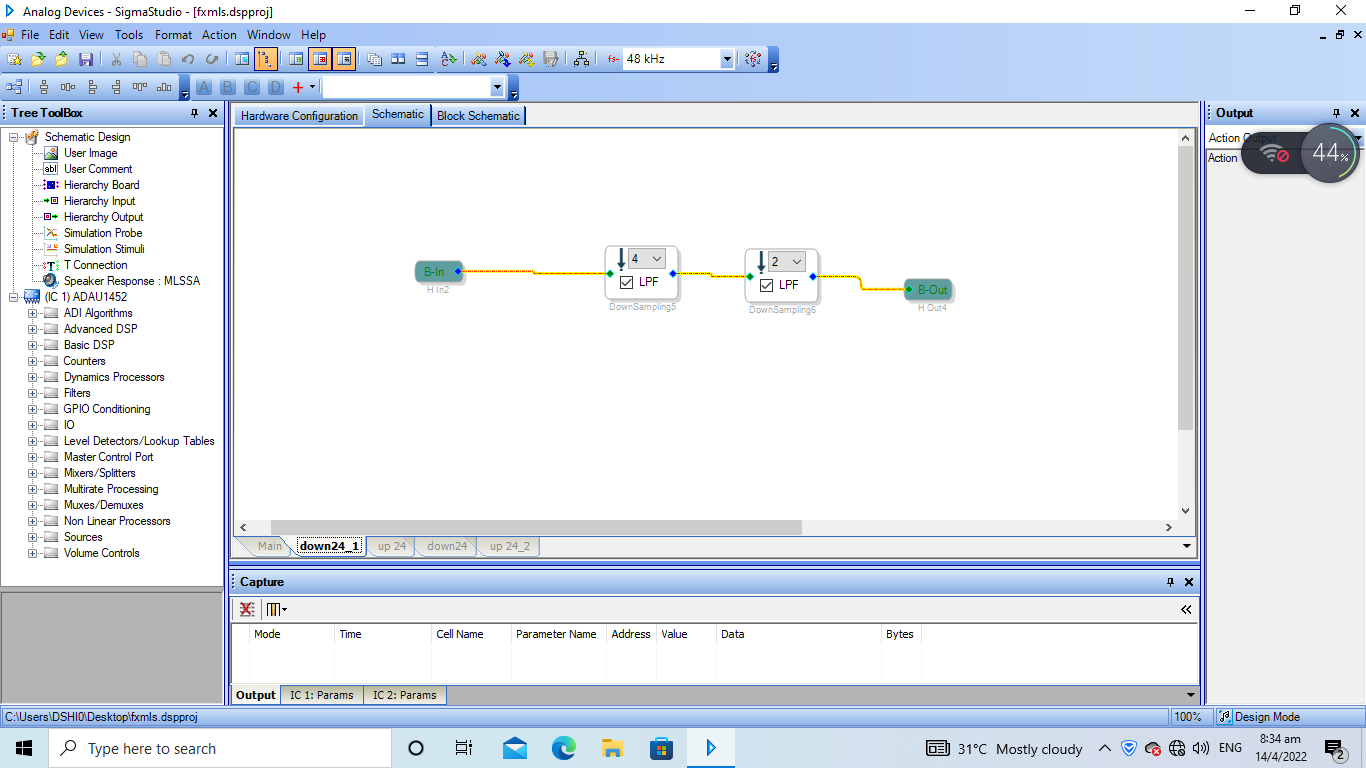}
  \caption{Down sampling function}
  \label{Fig29}
\end{figure}
\begin{figure}[htbp]
  \centering
    \includegraphics[width=13cm]{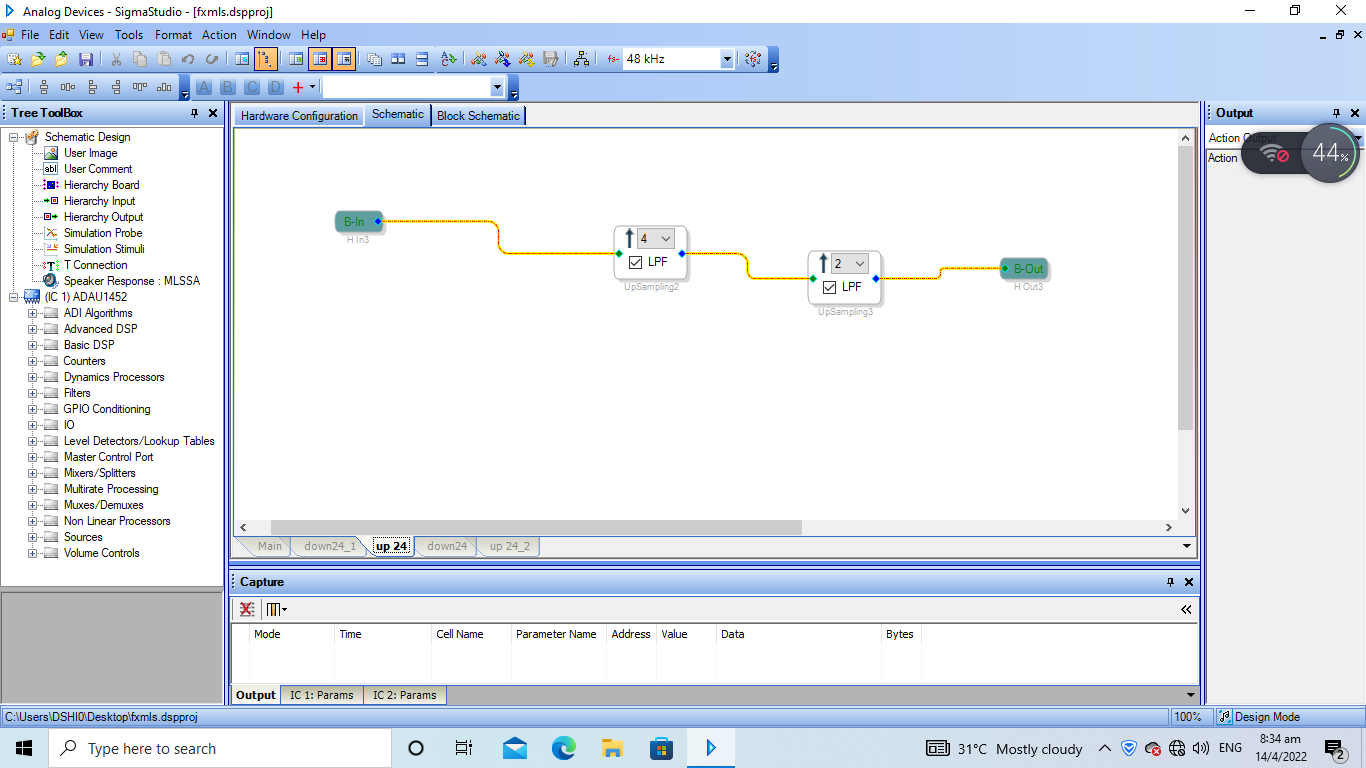}
  \caption{Up sampling function}
  \label{Fig30}
\end{figure}
The overall value of down-sampling or up-sampling (in terms of multiplication) is obtained by multiplying the values of two separate up-sampling and down-sampling processes. Additional low pass filters can be incorporated in order to get higher values. It is anticipated that modifying the sample rate has the potential to enhance noise compression.

\subsection{Experiment process}
The system is designed to address two distinct forms of noise. The first kind is subjected to filtration through a band-pass filter with a permissible frequency range of 300 Hz to 500 Hz. The second type corresponds to the actual noise produced by the coffee machine, and its spectrogram is depicted in Figure~\ref{Fig30}. 
\begin{figure}[htbp]
  \centering
    \includegraphics[width=13cm]{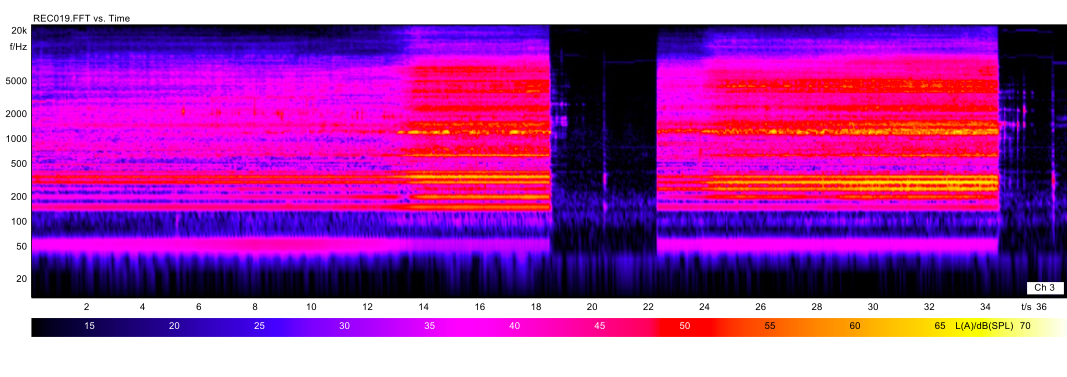}
  \caption{The spectrogram of coffee machine noise}
  \label{Fig31}
\end{figure}

The outcome of the preceding section is employed, wherein a modified FxLMS function block with a filter length of 128 and a training step size of $10^{-6}$ is implemented. Different sample rates are examined to determine the system characteristic that can effectively attenuate the widest range of noise frequencies.

The objective of the experimental procedure is to identify an initial system configuration that is approximately optimal. This involves establishing a fixed sampling rate for the program and adjusting the positions of the anti-noise speakers and error microphone. Next, it is necessary to reposition the microphone and speakers, as well as make appropriate adjustments to the sampling rate. Ultimately, the adjustment of the microphone and speakers' placement is necessary in order to complete the system design.

Initially, the sampling rate is set at a fixed value of 6 kHz. The determination of the position and angle of anti-noise sources involves doing numerous trials. Initially, the anti-noise speakers are placed adjacent to the noise source, while the error microphone is positioned at distances of 25 cm, 35cm, and 45cm from it. A noise compression of approximately 3dB occurs when the error microphone is positioned at distances of either 35cm or 45cm from the noise source. Subsequently, the anti-noise speakers are positioned adjacent to the test bench, with a separation of 10 cm, 20 cm, and 30 cm between each speaker and the noise source. The optimal outcome for reducing tone noise by 2dB is achieved when the positions are set at 10cm and 20 cm. Finally, the placement of the anti-noise speakers in opposition to the noise source is found to be impractical, as it aligns with the physical principles of wave propagation.

The results corresponding to the optimal system configuration are presented in Table~\ref{table7}.
\begin{table}[htbp]
\caption{Noise reduction performance of open space ANC system}\label{table7}
\centering
\begin{tabular}{|c|c|c|}
\hline
\multicolumn{1}{|l|}{\textbf{Sampling rate}} & \multicolumn{1}{l|}{\textbf{Filtered noise}} & \multicolumn{1}{l|}{\textbf{Actual noise}} \\ \hline
48khz/4                                      & 7dB, but the processing is slow.             & 3dB                                        \\ \hline
48khz/8                                      & 6dB                                          & 2$\sim$3dB                                 \\ \hline
48khz/12                                     & 6dB                                          & 1dB                                        \\ \hline
\end{tabular}
\end{table}
The performance of the sampling rate at 12kHz is exceptional; yet, the high sampling rate poses a threat due to its low processing and convergence speed.
Following the fine-tuning process, the anti-noise speakers have been positioned at a set distance of 5cm adjacent to the noise source. Consequently, both the noise wave and the anti-noise wave propagate in a parallel fashion. The error microphone is positioned at a distance of 40 cm in an open space environment. When the error microphone is moved within a radius of around 3cm, its displacement has a minimal impact on the final result.

\subsection{Experiment conclusion}
In summary, the ANC system that has been developed, in conjunction with the adapted FxLMS algorithm, demonstrates the capability to effectively attenuate noise levels in optimal circumstances characterized by minimal disturbances and straightforward sound propagation paths. This is evident in the results obtained from the duct experiment, where the sound propagation direction remains constant and extraneous noises are effectively blocked.

In the context of an open space environment, the system continues to demonstrate its effectiveness in efficiently mitigating the presence of filtered narrowband noise. The observed reduction in noise levels is substantial enough to be perceptible to individuals. Nevertheless, the system exhibits limited proficiency in dealing with unpredictable and fluctuating real-life noises due to the intricate nature of noise and the inherent constraints of the tools employed. The propagation of noise in open space occurs in multiple directions. However, our system is equipped with only one error microphone and two anti-noise speakers. It is possible that the system's performance could be improved by covering a greater range of directions. Additionally, there are hardware limitations that contribute to the system's shortcomings, such as delays in pre-amplifiers and limitations in the processing power of the chip.

The tests conducted aimed to validate the efficacy and underlying processes of the improved FXLMS algorithm. Furthermore, the optimal configurations of function block parameters, as well as the physical layouts of the platform and the test bench, are attained. Finally, suggestions are put forth to enhance the overall system functionality as well as the physical attributes of each individual component.  
\lhead{Potential Future Work}
\chapter{5. Potential Future Work}
The present experimental findings exhibit significant potential for enhancement. There is potential for the implementation of a more pronounced noise reduction impact that would be perceptible to human individuals. Subsequently, it becomes feasible to advocate for the application of this technology in a commercial context. Hence, based on the empirical observations made during the experiment and the theoretical frameworks outlined in pertinent scholarly literature, two recommendations are put out for future investigation.

At the outset, there exists the potential to use or create exceedingly efficient electrical components, such as amplifiers and loudspeakers, as these serve as the primary contributors to delay~\cite{self2013audio}. The experiment centers around the study of white noise in an unobstructed environment. The noise source exhibits persistent temporal variability, and furthermore, white noise is characterized by its inherent lack of predictability. Consequently, any significant delay in the experiment is deemed unsatisfactory.
The experimental findings indicate that the system demonstrates efficiency in mitigating noise within a specific range of frequencies~\cite{thi1993delayless}. One plausible suggestion is to enhance the effectiveness of active noise control (ANC) by incorporating additional reference signals and control sources. This approach involves combining numerous ANC systems into a unified framework, where each system is dedicated to addressing a certain frequency range of noise. In order to achieve the desired outcome, the utilization of bandpass filters is essential for obtaining the necessary reference signals for each subsystem. Additionally, a modified algorithm will be required to counteract the impact of other control sources at each error microphone location. One significant limitation of the concept is the intricate nature of the system, both in terms of computational and physical aspects. As active noise control (ANC) systems continue to evolve towards smaller and more portable designs to cater to various applications such household appliances and earpieces, the inclusion of sensors and adaptive filters incurs a substantial expense.
\addcontentsline{toc}{chapter}{References} 
\renewcommand{\baselinestretch}{1.0}
\printbibliography[title={References}]

@article{lam2021ten,
  title={Ten questions concerning active noise control in the built environment},
  author={Lam, Bhan and Gan, Woon-Seng and Shi, DongYuan and Nishimura, Masaharu and Elliott, Stephen},
  journal={Building and Environment},
  volume={200},
  pages={107928},
  year={2021},
  publisher={Elsevier}
}

@article{elliott1993active,
  title={Active noise control},
  author={Elliott, Stephen J and Nelson, Philip Arthur},
  journal={IEEE signal processing magazine},
  volume={10},
  number={4},
  pages={12--35},
  year={1993},
  publisher={IEEE}
}

@article{shi2023active,
  title={Active Noise Control in The New Century: The Role and Prospect of Signal Processing},
  author={Shi, Dongyuan and Lam, Bhan and Gan, Woon-Seng and Cheer, Jordan},
  journal={arXiv preprint arXiv:2306.01425},
  year={2023}
}

@article{kuo1996design,
  title={Design of active noise control systems with the TMS320 family},
  author={Kuo, Sen M and Panahi, Issa and Chung, Kai M and Horner, Tom and Nadeski, Mark and Chyan, Jason},
  journal={Texas Instruments},
  year={1996}
}

@inproceedings{shi2016comparison,
  title={Comparison of different development kits and its suitability in signal processing education},
  author={Shi, Dong Yuan and Gan, Woon-Seng},
  booktitle={2016 IEEE International Conference on Acoustics, Speech and Signal Processing (ICASSP)},
  pages={6280--6284},
  year={2016},
  organization={IEEE}
}

@article{burgess1981active,
  title={Active adaptive sound control in a duct: A computer simulation},
  author={Burgess, John C},
  journal={The Journal of the Acoustical Society of America},
  volume={70},
  number={3},
  pages={715--726},
  year={1981},
  publisher={Acoustical Society of America}
}

@misc{moazzam2014performance,
  title={Performance evaluation of different active noise control (ANC) algorithms for attenuating noise in a duct},
  author={Moazzam, Muhammad and Rabbani, Muhammad Shoaib},
  year={2014}
}

@inproceedings{shi2017effect,
  title={Effect of the audio amplifier's distortion on feedforward active noise control},
  author={Shi, Dongyuan and Shi, Chuang and Gan, Woon-Seng},
  booktitle={2017 Asia-Pacific Signal and Information Processing Association Annual Summit and Conference (APSIPA ASC)},
  pages={469--473},
  year={2017},
  organization={IEEE}
}

@article{shi2021optimal,
  title={Optimal output-constrained active noise control based on inverse adaptive modeling leak factor estimate},
  author={Shi, Dongyuan and Gan, Woon-Seng and Lam, Bhan and Wen, Shulin and Shen, Xiaoyi},
  journal={IEEE/ACM Transactions on Audio, Speech, and Language Processing},
  volume={29},
  pages={1256--1269},
  year={2021},
  publisher={IEEE}
}

@article{qiu2002waveform,
  title={A waveform synthesis algorithm for active control of transformer noise: implementation},
  author={Qiu, Xiaojun and Li, Xun and Ai, Yanting and Hansen, Colin H},
  journal={Applied Acoustics},
  volume={63},
  number={5},
  pages={467--479},
  year={2002},
  publisher={Elsevier}
}

@article{fatima2012noise,
  title={Noise control of home appliances—The green way},
  author={Fatima, S and Mohanty, AR},
  journal={Noise \& Vibration Worldwide},
  volume={43},
  number={7},
  pages={26--34},
  year={2012},
  publisher={SAGE Publications Sage UK: London, England}
}

@article{elliot1990flight,
  title={In-flight experiments on the active control of propeller-induced cabin noise},
  author={Elliot, Stephen J and Nelson, Philip A and Stothers, Ian M and Boucher, Christopher C},
  journal={Journal of Sound and Vibration},
  volume={140},
  number={2},
  pages={219--238},
  year={1990},
  publisher={Elsevier}
}

@article{lam2020active,
  title={Active control of broadband sound through the open aperture of a full-sized domestic window},
  author={Lam, Bhan and Shi, Dongyuan and Gan, Woon-Seng and Elliott, Stephen J and Nishimura, Masaharu},
  journal={Scientific reports},
  volume={10},
  number={1},
  pages={1--7},
  year={2020},
  publisher={Nature Publishing Group}
}

@article{lam2018active,
  title={Active control of sound through full-sized open windows},
  author={Lam, Bhan and Shi, Chuang and Shi, Dongyuan and Gan, Woon-Seng},
  journal={Building and Environment},
  volume={141},
  pages={16--27},
  year={2018},
  publisher={Elsevier}
}

@inproceedings{shi2016open,
  title={Open loop active control of noise through open windows},
  author={Shi, Chuang and Murao, Tatsuya and Shi, Dongyuan and Lam, Bhan and Gan, Woon-Seng},
  booktitle={Proceedings of Meetings on Acoustics},
  volume={29},
  number={1},
  year={2016},
  organization={AIP Publishing}
}

@article{shen2022adaptive,
  title={Adaptive-gain algorithm on the fixed filters applied for active noise control headphone},
  author={Shen, Xiaoyi and Shi, Dongyuan and Gan, Woon-Seng and Peksi, Santi},
  journal={Mechanical Systems and Signal Processing},
  volume={169},
  pages={108641},
  year={2022},
  publisher={Elsevier}
}

@article{shen2022multi,
  title={Multi-channel wireless hybrid active noise control with fixed-adaptive control selection},
  author={Shen, Xiaoyi and Gan, Woon-Seng and Shi, Dongyuan},
  journal={Journal of Sound and Vibration},
  volume={541},
  pages={117300},
  year={2022},
  publisher={Elsevier}
}

@article{shen2021alternative,
  title={Alternative switching hybrid ANC},
  author={Shen, Xiaoyi and Gan, Woon-Seng and Shi, Dongyuan},
  journal={Applied Acoustics},
  volume={173},
  pages={107712},
  year={2021},
  publisher={Elsevier}
}

@inproceedings{shen2023implementations,
  title={Implementations of wireless active noise control in the headrest},
  author={Shen, Xiaoyi and Shi, Dongyuan and Peksi, Santi and Gan, Woon-Seng},
  booktitle={INTER-NOISE and NOISE-CON Congress and Conference Proceedings},
  volume={265},
  number={4},
  pages={3445--3455},
  year={2023},
  organization={Institute of Noise Control Engineering}
}

@inproceedings{shen2021wireless,
  title={A wireless reference active noise control headphone using coherence based selection technique},
  author={Shen, Xiaoyi and Shi, Dongyuan and Gan, Woon-Seng},
  booktitle={ICASSP 2021-2021 IEEE International Conference on Acoustics, Speech and Signal Processing (ICASSP)},
  pages={7983--7987},
  year={2021},
  organization={IEEE}
}

@article{lee2022compact,
  title={Compact hybrid noise control system: Anc system equipped with circular noise barrier using theoretically calculated control filter},
  author={Lee, Sanghyeon and Park, Youngjin},
  journal={Applied Acoustics},
  volume={188},
  pages={108472},
  year={2022},
  publisher={Elsevier}
}

@article{shi2020algorithms,
  title={Algorithms and implementations to overcome practical issues in active noise control systems},
  author={Shi, Dongyuan},
  year={2020},
  publisher={Nanyang Technological University}
}

@book{hansen1999understanding,
  title={Understanding active noise cancellation},
  author={Hansen, Colin H},
  year={1999},
  publisher={CRC Press}
}

@book{kuo1996active,
  title={Active noise control systems},
  author={Kuo, Sen M and Morgan, Dennis R},
  volume={4},
  year={1996},
  publisher={New York: Wiley}
}

@inproceedings{shen2021implementation,
  title={Implementation of coherence-based-selection multi-channel wireless active noise control in headphone},
  author={Shen, Xiaoyi and Shi, Dongyuan and Gan, Woon-Seng and Peksi, Santi},
  booktitle={INTER-NOISE and NOISE-CON Congress and Conference Proceedings},
  volume={263},
  number={4},
  pages={1945--1953},
  year={2021},
  organization={Institute of Noise Control Engineering}
}

@inproceedings{shi2017multiple,
  title={Multiple parallel branch with folding architecture for multichannel filtered-x least mean square algorithm},
  author={Shi, Dongyuan and He, Jianjun and Shi, Chuang and Murao, Tatsuya and Gan, Woon-Seng},
  booktitle={2017 IEEE International Conference on Acoustics, Speech and Signal Processing (ICASSP)},
  pages={1188--1192},
  year={2017},
  organization={IEEE}
}

@inproceedings{shi2017understanding,
  title={Understanding multiple-input multiple-output active noise control from a perspective of sampling and reconstruction},
  author={Shi, Chuang and Li, Huiyong and Shi, Dongyuan and Lam, Bhan and Gan, Woon-Seng},
  booktitle={2017 Asia-Pacific Signal and Information Processing Association Annual Summit and Conference (APSIPA ASC)},
  pages={124--129},
  year={2017},
  organization={IEEE}
}

@article{he2019exploiting,
  title={Exploiting the underdetermined system in multichannel active noise control for open windows},
  author={He, Jianjun and Lam, Bhan and Shi, Dongyuan and Gan, Woon Seng},
  journal={Applied Sciences},
  volume={9},
  number={3},
  pages={390},
  year={2019},
  publisher={MDPI}
}

@inproceedings{shi2017algorithms,
  title={On algorithms and implementations of a 4-channel active noise canceling window},
  author={Shi, Chuang and Jiang, Nan and Li, Huiyong and Shi, Dongyuan and Gan, Woon-Seng},
  booktitle={2017 International Symposium on Intelligent Signal Processing and Communication Systems (ISPACS)},
  pages={217--221},
  year={2017},
  organization={IEEE}
}

@inproceedings{hasegawa2018window,
  title={Window active noise control system with virtual sensing technique},
  author={Hasegawa, Rina and Shi, Dongyuan and Kajikawa, Yoshinobu and Gan, Woon-Seng},
  booktitle={INTER-NOISE and NOISE-CON Congress and Conference Proceedings},
  volume={258},
  number={1},
  pages={6004--6012},
  year={2018},
  organization={Institute of Noise Control Engineering}
}

@article{widrow1975adaptive,
  title={Adaptive noise cancelling: Principles and applications},
  author={Widrow, Bernard and Glover, John R and McCool, John M and Kaunitz, John and Williams, Charles S and Hearn, Robert H and Zeidler, James R and Dong, JR Eugene and Goodlin, Robert C},
  journal={Proceedings of the IEEE},
  volume={63},
  number={12},
  pages={1692--1716},
  year={1975},
  publisher={IEEE}
}

@book{nelson1991active,
  title={Active control of sound},
  author={Nelson, Philip Arthur and Elliott, Stephen J},
  year={1991},
  publisher={Academic press}
}

@article{george2013advances,
  title={Advances in active noise control: A survey, with emphasis on recent nonlinear techniques},
  author={George, Nithin V and Panda, Ganapati},
  journal={Signal processing},
  volume={93},
  number={2},
  pages={363--377},
  year={2013},
  publisher={Elsevier}
}

@inproceedings{shi2016systolic,
  title={A systolic FxLMS structure for implementation of feedforward active noise control on FPGA},
  author={Shi, Dongyuan and Shi, Chuang and Gan, Woon-Seng},
  booktitle={2016 Asia-Pacific Signal and Information Processing Association Annual Summit and Conference (APSIPA)},
  pages={1--6},
  year={2016},
  organization={IEEE}
}

@article{shi2019two,
  title={Two-gradient direction FXLMS: An adaptive active noise control algorithm with output constraint},
  author={Shi, DongYuan and Gan, Woon-Seng and Lam, Bhan and Shi, Chuang},
  journal={Mechanical Systems and Signal Processing},
  volume={116},
  pages={651--667},
  year={2019},
  publisher={Elsevier}
}

@article{shi2019optimal,
  title={Optimal leak factor selection for the output-constrained leaky filtered-input least mean square algorithm},
  author={Shi, Dongyuan and Lam, Bhan and Gan, Woon-Seng and Wen, Shulin},
  journal={IEEE Signal Processing Letters},
  volume={26},
  number={5},
  pages={670--674},
  year={2019},
  publisher={IEEE}
}

@article{shi2020feedforward,
  title={Feedforward multichannel virtual-sensing active control of noise through an aperture: Analysis on causality and sensor-actuator constraints},
  author={Shi, Dongyuan and Gan, Woon-Seng and Lam, Bhan and Hasegawa, Rina and Kajikawa, Yoshinobu},
  journal={The Journal of the Acoustical Society of America},
  volume={147},
  number={1},
  pages={32--48},
  year={2020},
  publisher={AIP Publishing}
}

@article{lam2023anti,
  title={Anti-noise window: Subjective perception of active noise reduction and effect of informational masking},
  author={Lam, Bhan and Lim, Kelvin Chee Quan and Ooi, Kenneth and Ong, Zhen-Ting and Shi, Dongyuan and Gan, Woon-Seng},
  journal={Sustainable Cities and Society},
  volume={97},
  pages={104763},
  year={2023},
  publisher={Elsevier}
}

@article{elliott1987multiple,
  title={A multiple error LMS algorithm and its application to the active control of sound and vibration},
  author={Elliott, Stephen and Stothers, IANM and Nelson, Philip},
  journal={IEEE Transactions on Acoustics, Speech, and Signal Processing},
  volume={35},
  number={10},
  pages={1423--1434},
  year={1987},
  publisher={IEEE}
}

@article{kuo1999active,
  title={Active noise control: a tutorial review},
  author={Kuo, Sen M and Morgan, Dennis R},
  journal={Proceedings of the IEEE},
  volume={87},
  number={6},
  pages={943--973},
  year={1999},
  publisher={IEEE}
}

@article{shi2019practical,
  title={Practical implementation of multichannel filtered-x least mean square algorithm based on the multiple-parallel-branch with folding architecture for large-scale active noise control},
  author={Shi, Dongyuan and Gan, Woon-Seng and He, Jianjun and Lam, Bhan},
  journal={IEEE Transactions on Very Large Scale Integration (VLSI) Systems},
  volume={28},
  number={4},
  pages={940--953},
  year={2019},
  publisher={IEEE}
}

@inproceedings{shi2020multichannel,
  title={Multichannel active noise control with spatial derivative constraints to enlarge the quiet zone},
  author={Shi, Dongyuan and Lam, Bhan and Wen, Shulin and Gan, Woon-Seng},
  booktitle={ICASSP 2020-2020 IEEE International Conference on Acoustics, Speech and Signal Processing (ICASSP)},
  pages={8419--8423},
  year={2020},
  organization={IEEE}
}

@inproceedings{shi2020active,
  title={Active noise control based on the momentum multichannel normalized filtered-x least mean square algorithm},
  author={Shi, Dongyuan and Gan, Woon-Seng and Lam, Bhan and Wen, Shulin and Shen, Xiaoyi},
  booktitle={INTER-NOISE and NOISE-CON Congress and Conference Proceedings},
  volume={261},
  number={6},
  pages={709--719},
  year={2020},
  organization={Institute of Noise Control Engineering}
}

@article{shi2021block,
  title={Block coordinate descent based algorithm for computational complexity reduction in multichannel active noise control system},
  author={Shi, Dongyuan and Lam, Bhan and Gan, Woon-Seng and Wen, Shulin},
  journal={Mechanical Systems and Signal Processing},
  volume={151},
  pages={107346},
  year={2021},
  publisher={Elsevier}
}

@article{shi2023multichannel,
  title={Multichannel two-gradient direction filtered reference least mean square algorithm for output-constrained multichannel active noise control},
  author={Shi, Dongyuan and Lam, Bhan and Shen, Xiaoyi and Gan, Woon-Seng},
  journal={Signal Processing},
  volume={207},
  pages={108938},
  year={2023},
  publisher={Elsevier}
}

@article{shi2023computation,
  title={Computation-efficient solution for fully-connected active noise control window: Analysis and implementation of multichannel adjoint least mean square algorithm},
  author={Shi, Dongyuan and Lam, Bhan and Ji, Junwei and Shen, Xiaoyi and Lai, Chung Kwan and Gan, Woon-Seng},
  journal={Mechanical Systems and Signal Processing},
  volume={199},
  pages={110444},
  year={2023},
  publisher={Elsevier}
}

@inproceedings{shen2023momentum,
  title={A Momentum Two-Gradient Direction Algorithm with Variable Step Size Applied to Solve Practical Output Constraint Issue for Active Noise Control},
  author={Shen, Xiaoyi and Shi, Dongyuan and Luo, Zhengding and Ji, Junwei and Gan, Woon-Seng},
  booktitle={ICASSP 2023-2023 IEEE International Conference on Acoustics, Speech and Signal Processing (ICASSP)},
  pages={1--5},
  year={2023},
  organization={IEEE}
}

@article{shi2021comb,
  title={Comb-partitioned frequency-domain constraint adaptive algorithm for active noise control},
  author={Shi, Dongyuan and Gan, Woon-Seng and Lam, Bhan and Shen, Xiaoyi},
  journal={Signal Processing},
  volume={188},
  pages={108222},
  year={2021},
  publisher={Elsevier}
}

@article{akhtar2009improving,
  title={Improving performance of FxLMS algorithm for active noise control of impulsive noise},
  author={Akhtar, Muhammad Tahir and Mitsuhashi, Wataru},
  journal={Journal of Sound and Vibration},
  volume={327},
  number={3-5},
  pages={647--656},
  year={2009},
  publisher={Elsevier}
}

@inproceedings{akhtar2004modified,
  title={Modified-filtered-x LMS algorithm based active noise control systems with improved online secondary-path modeling},
  author={Akhtar, M Tahir and Abe, Masahide and Kawamata, Masayuki},
  booktitle={The 2004 47th Midwest Symposium on Circuits and Systems, 2004. MWSCAS'04.},
  volume={1},
  pages={I--13},
  year={2004},
  organization={IEEE}
}

@article{lai2023mov,
  title={MOV-Modified-FxLMS algorithm with Variable Penalty Factor in a Practical Power Output Constrained Active Control System},
  author={Lai, Chung Kwan and Shi, Dongyuan and Lam, Bhan and Gan, Woon-Seng},
  journal={IEEE Signal Processing Letters},
  year={2023},
  publisher={IEEE}
}

@article{luo2022implementation,
  title={Implementation of multi-channel active noise control based on back-propagation mechanism},
  author={Luo, Zhengding and Shi, Dongyuan and Ji, Junwei and Gan, Woon-seng},
  journal={arXiv preprint arXiv:2208.08086},
  year={2022}
}

@article{shi2022selective,
  title={Selective fixed-filter active noise control based on convolutional neural network},
  author={Shi, Dongyuan and Lam, Bhan and Ooi, Kenneth and Shen, Xiaoyi and Gan, Woon-Seng},
  journal={Signal Processing},
  volume={190},
  pages={108317},
  year={2022},
  publisher={Elsevier}
}

@article{luo2022hybrid,
  title={A hybrid sfanc-fxnlms algorithm for active noise control based on deep learning},
  author={Luo, Zhengding and Shi, Dongyuan and Gan, Woon-Seng},
  journal={IEEE Signal Processing Letters},
  volume={29},
  pages={1102--1106},
  year={2022},
  publisher={IEEE}
}

@inproceedings{luo2023performance,
  title={Performance Evaluation of Selective Fixed-filter Active Noise Control based on Different Convolutional Neural Networks},
  author={Luo, Zhengding and Shi, Dongyuan and Gan, Woon-Seng and Huang, Qirui and Zhang, Libin},
  booktitle={INTER-NOISE and NOISE-CON Congress and Conference Proceedings},
  volume={265},
  number={6},
  pages={1615--1622},
  year={2023},
  organization={Institute of Noise Control Engineering}
}

@article{shi2023transferable,
  title={Transferable latent of cnn-based selective fixed-filter active noise control},
  author={Shi, Dongyuan and Gan, Woon-Seng and Lam, Bhan and Luo, Zhengding and Shen, Xiaoyi},
  journal={IEEE/ACM Transactions on Audio, Speech, and Language Processing},
  year={2023},
  publisher={IEEE}
}

@inproceedings{luo2023deep,
  title={Deep Generative Fixed-Filter Active Noise Control},
  author={Luo, Zhengding and Shi, Dongyuan and Shen, Xiaoyi and Ji, Junwei and Gan, Woon-Seng},
  booktitle={ICASSP 2023-2023 IEEE International Conference on Acoustics, Speech and Signal Processing (ICASSP)},
  pages={1--5},
  year={2023},
  organization={IEEE}
}

@article{shi2021fast,
  title={Fast adaptive active noise control based on modified model-agnostic meta-learning algorithm},
  author={Shi, Dongyuan and Gan, Woon-Seng and Lam, Bhan and Ooi, Kenneth},
  journal={IEEE Signal Processing Letters},
  volume={28},
  pages={593--597},
  year={2021},
  publisher={IEEE}
}

@inproceedings{li2013frequency,
  title={Frequency estimation on power system using recursive-least-squares approach},
  author={Li, Liangliang and Xia, Wei and Shi, Dongyuan and Li, Jianzhuang},
  booktitle={Proceedings of the 2012 International Conference on Information Technology and Software Engineering: Information Technology \& Computing Intelligence},
  pages={11--18},
  year={2013},
  organization={Springer}
}

@book{farhang2013adaptive,
  title={Adaptive filters: theory and applications},
  author={Farhang-Boroujeny, Behrouz},
  year={2013},
  publisher={John Wiley \& Sons}
}

@book{haykin2002adaptive,
  title={Adaptive filter theory},
  author={Haykin, Simon S},
  year={2002},
  publisher={Pearson Education India}
}

@misc{shi2016adaptive,
  title={Adaptive filtering method and system based on error sub-band},
  author={Shi, Dongyuan and He, Dongmei and Cai, Meng},
  year={2016},
  month=aug # "~16",
  publisher={Google Patents},
  note={US Patent 9,419,826}
}

@inproceedings{shi2018novel,
  title={A novel selective active noise control algorithm to overcome practical implementation issue},
  author={Shi, Dong Yuan and Lam, Bhan and Gan, Woon-Seng},
  booktitle={2018 IEEE International Conference on Acoustics, Speech and Signal Processing (ICASSP)},
  pages={1130--1134},
  year={2018},
  organization={IEEE}
}

@article{wen2020using,
  title={Using empirical wavelet transform to speed up selective filtered active noise control system},
  author={Wen, Shulin and Gan, Woon-Seng and Shi, Dongyuan},
  journal={The Journal of the Acoustical Society of America},
  volume={147},
  number={5},
  pages={3490--3501},
  year={2020},
  publisher={AIP Publishing}
}

@inproceedings{shi2019analysis,
  title={Analysis of multichannel virtual sensing active noise control to overcome spatial correlation and causality constraints},
  author={Shi, Dongyuan and Lam, Bhan and Gan, Woon-seng},
  booktitle={ICASSP 2019-2019 IEEE International Conference on Acoustics, Speech and Signal Processing (ICASSP)},
  pages={8499--8503},
  year={2019},
  organization={IEEE}
}

@inproceedings{lai2023real,
  title={Real-time modelling of observation filter in the Remote Microphone Technique for an Active Noise Control application},
  author={Lai, Chung Kwan and Lam, Bhan and Shi, Dongyuan and Gan, Woon-Seng},
  booktitle={ICASSP 2023-2023 IEEE International Conference on Acoustics, Speech and Signal Processing (ICASSP)},
  pages={1--5},
  year={2023},
  organization={IEEE}
}

@article{gan2023practical,
  title={Practical Active Noise Control: Restriction of Maximum Output Power},
  author={Gan, Woon-Seng and Shi, Dongyuan and Shen, Xiaoyi},
  journal={arXiv preprint arXiv:2307.10913},
  year={2023}
}

@inproceedings{lai2023robust,
  title={Robust estimation of open aperture active control systems using virtual sensing},
  author={Lai, Chung Kwan and Tey, Jing Sheng and Shi, Dongyuan and Gan, Woon-Seng},
  booktitle={INTER-NOISE and NOISE-CON Congress and Conference Proceedings},
  volume={265},
  number={4},
  pages={3397--3407},
  year={2023},
  organization={Institute of Noise Control Engineering}
}

@online{ADAU1452,
  author = {Analog Devices},
  title = {SigmaDSP digital audio processor- ADAU145X},
  year = 2014,
  url = {https://www.analog.com/media/en/technical-documentation/data-sheets/ADAU1452_1451_1450.pdf},
  urldate = {2014-09-30}
}

@online{ADAU1452_EVM,
  author = {Analog Devices},
  title = {UG-1662: EVAL-ADAU1452REVBZ User Guide },
  year = 2019,
  url = {chrome-extension://efaidnbmnnnibpcajpcglclefindmkaj/https://www.analog.com/media/en/technical-documentation/user-guides/EVAL-ADAU1452REVBZ-UG-1662.pdf},
  urldate = {2019-09-30}
}

@book{self2013audio,
  title={Audio power amplifier design},
  author={Self, Douglas},
  year={2013},
  publisher={Taylor \& Francis}
}

@inproceedings{thi1993delayless,
  title={Delayless subband active noise control},
  author={Thi, James and Morgan, Dennis R},
  booktitle={1993 IEEE International Conference on Acoustics, Speech, and Signal Processing},
  volume={1},
  pages={181--184},
  year={1993},
  organization={IEEE}
}

  \clearpage
  \pagenumbering{arabic}%
  \renewcommand{\thepage}{R-\arabic{page}}
\lhead{}

\newpage
\appendix
\renewcommand{\chapname}{Appendix}
\pretocmd{\chapter}{%
  \clearpage
  \pagenumbering{arabic}%
  \renewcommand*{\thepage}{\thechapter-\arabic{page}}%
}{}{}
\rhead{Appendix}

\chapter{Appendix Figures}
\section{Matlab code of the LMS algorithm}
\begin{figure}[htbp]
    \centering
    \includegraphics[width=14cm]{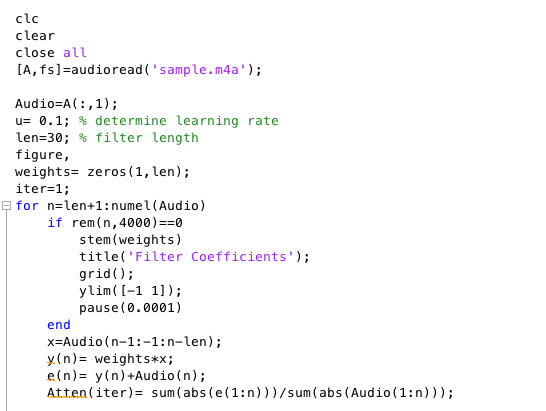}
    \caption{The LMS algorithm (Part I)}
    \label{fig:e32}
\end{figure}
\begin{figure}[htbp]
    \centering
    \includegraphics[width=14cm]{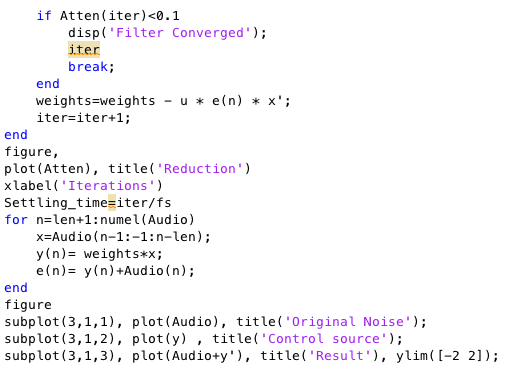}
    \caption{The LMS algorithm (Part II)}
    \label{fig:e33}
\end{figure}

\section{Matlab code of the FxLMS algorithm}
\begin{figure}[htbp]
    \centering
    \includegraphics[width=14cm]{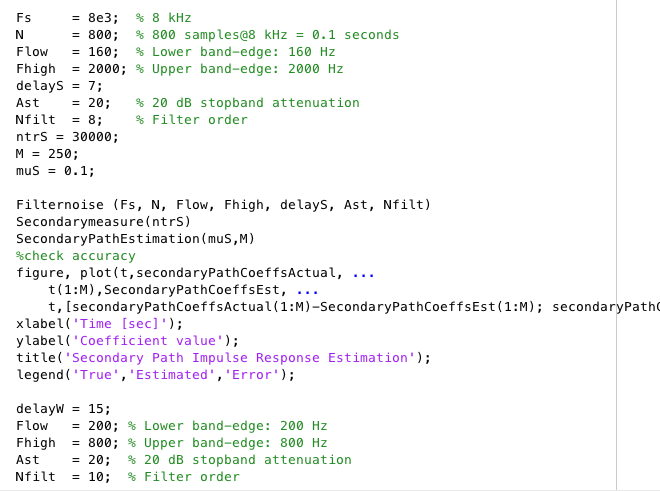}
    \caption{The FxLMS algorithm (Part I)}
    \label{fig:e34}
\end{figure}
\begin{figure}[htbp]
    \centering
    \includegraphics[width=14cm]{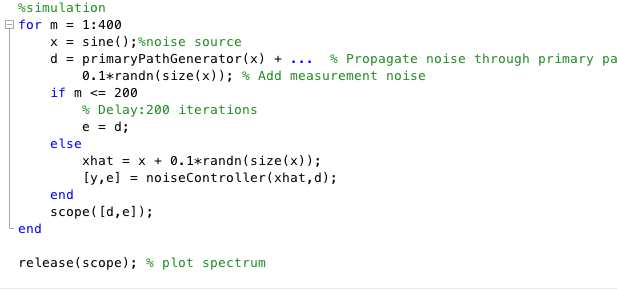}
    \caption{The FxLMS algorithm (Part II)}
    \label{fig:e35}
\end{figure}
\begin{figure}[htbp]
    \centering
    \includegraphics[width=14cm]{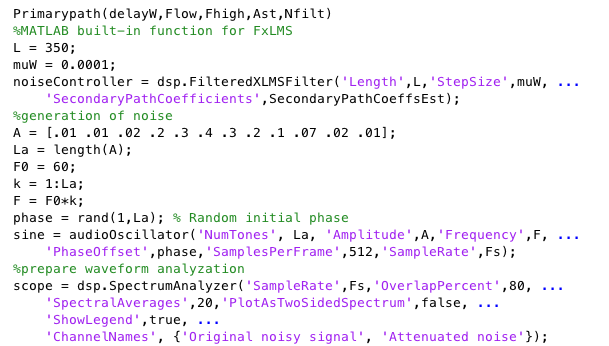}
    \caption{The FxLMS algorithm (Part III)}
    \label{fig:e36}
\end{figure}

\end{document}